\documentclass[12pt]{iopart}
\usepackage{iopams}
\usepackage{dsfont}
\usepackage{xcolor}
\usepackage{amssymb}
\usepackage[utf8]{inputenc}
\usepackage{graphicx}
\usepackage{bm,bbm}

\DeclareUnicodeCharacter{03B1}{$\alpha$}
\newcommand{\ee}[1]{\mathrm{e}^{#1}}
\newcommand{\erfi}{\mathrm{erfi}}
\newcommand{\tkappa}{\tilde{K}}

\begin{document}
\title[The
  Gaussian Network Model]{Time- and ensemble-average statistical mechanics of the
  Gaussian Network Model}
\author{Alessio Lapolla, Maximilian Vossel and Alja\v{z} Godec}
\address{Mathematical bioPhysics group, Max Planck Institute for
  Biophysical Chemistry, Am Fassberg 11, G\"{o}ttingen 37077, Germany}
\ead{agodec@mpibpc.mpg.de}
\vspace{10pt}

\begin{abstract}
We present analytical results (up to a numerical diagonalization of a
real symmetric matrix) for a set of time- and
ensemble-average physical observables in the non-Hookean Gaussian Network 
Model (GNM) -- a generalization of the Rouse model to elastic networks with
links with a certain degree of 
extensional and rotational stiffness. We focus on a set of
coarse-grained observables that may be of interest in the analysis of
GNM in the context of internal motions in proteins and mechanical frames in contact with a
heat bath. A C++ computer code is made available that implements all analytical
results. 
\end{abstract}
\submitto{\jpa}

\section{Introduction}
Proteins utilize their unique
dynamic character encoded in internal motions to execute a biological function \cite{Kern}. These
motions span fs to s time-scales and their study thus requires a multitude of
experimental and/or computational methods \cite{Kern}. The most
detailed, atomically resolved information about these
motions comes --  with a grain of salt because of an underlying
approximate, empirical potential energy function -- from Molecular Dynamics (MD)
simulations \cite{Berendsen,Shaw}. However, even if the
state-of-the-art hardware and highly parallel algorithms allow to
reach ms time-scales \cite{Shaw_2} a substantial time-scale gap remains. In
addition, the sheer amount of detail in such \emph{tour de force} simulations
\cite{Shaw_2} often poses a challenge if one aims at extracting
minimal, ``leading order'' physical principles underlying protein
internal motions.  Moreover, physical or even topological properties alone
may accurately predict selected features of protein dynamics \cite{Baiesi,Seno}.

To describe internal motions in proteins on an effective,
coarse-grained level disregarding chemical details Tirion introduced
the so-called Elastic Network Model (ENM)~\cite{tirion_large_1996}
akin to the seminal works of
Rouse~\cite{rouse_theory_1953} and
Flory~\cite{flory_p_j_statistical_1976} in polymer physics.  The basic idea underlying ENMs is an elastic network
 connecting those residues, more precisely the respective C$\alpha$
 atoms, that lie within a cutoff distance typically chosen in the range
 $7-16$\,\AA. 
Subsequent
works considered various alternative models, e.g. so-called Gaussian Network Model (GNM)~\cite{bahar_direct_1997,
  haliloglu_gaussian_1997} and the Anisotropic Network Model
(ANM)~\cite{doruker_dynamics_2000, atilgan_anisotropy_2001}. 

Up do date elastic network models in various forms have
have been successfully applied (and extended) to refine
NMR-~\cite{delarue_use_2004} and X-ray crystallography-derived protein 
structures~\cite{schroder_combining_2007}, derive NMR-structural order
parameters~\cite{ming_reorientational_2006}, investigate structural
 correlations~\cite{tang_critical_2017}, function \cite{hamacher_dependency_2006,rader_identification_2004,
  keskin_relating_2002, weng_conformational_2008}, conformational transitions~\cite{putz_elastic_2017}, and allosteric
 effects~\cite{zheng_identification_2005} in proteins, and to identify
 and decompose protein domains~\cite{kundu_automatic_2004}. Further
applications involve 
improving Molecular
Dynamics simulations~\cite{zhang_molecular_2003}, the study of protein
evolution~\cite{hamacher_relating_2008}, investigations of smart
polymers~\cite{zhang_novel_2015, zhang_optimized_2016},
viruses~\cite{tama_mechanism_2002}, membrane
channels~\cite{shrivastava_common_2006, bocquet_x-ray_2009}, 
and nucleic acids~\cite{pinamonti_elastic_2015}, as well as the prediction of  rupture points in single-molecule 
pulling experiments~\cite{sulkowska_predicting_2008},

Most of these works rely on ``standard'' Normal Mode Analysis
(NMA)~\cite{goldstein_classical_2002, bahar_normal_2010}, i.e. on
spectral characteristics of the underlying mechanical vibration spectrum. In the particular context of proteins NMA has been used predominantly
 to identify the large-scale collective motions encoded in the
 eigenvector corresponding to the principal eigenvalue of the
 Hessian. Notably, the low-frequency modes are quite insensitive
 to the precise value of the cutoff distance~\cite{Nicolay_Functional_2006}.

Here we go beyond and present analytical results for time- and ensemble-average
characteristics of internal ``reaction coordinates'' in GNM in contact
with a heat bath at a finite temperature. More precisely, we 
consider the non-Markovian dynamics of internal distances at
equilibrium. Our results may be relevant for interpreting
single-molecule spectroscopy data or Molecular Dynamics simulations.

\section{The Gaussian Network Model}
The Rouse model ~\cite{rouse_theory_1953} is one of the earliest
``elastic network'' models of flexible linear polymers (later
on extended to more general network structures \cite{flory_p_j_statistical_1976}). It neglects
excluded volume effects and hydrodynamic interactions. 
Within this theoretical framework beads are connected by ideal,
Hookean springs with vanishing resting length (i.e. at $T=0$ the beads'
positions would coincide). The strength of the springs is
proportional to the temperature $T$ of the heat bath. The model does
not accurately capture the features of molecules with a non-negligibly
internal rigidity.

ENMs~\cite{tirion_large_1996} extend these core ideas by including
a non-zero resting length, i.e.  at $T=0$ the residues are assumed to have distinct
positions that are fixed in space. This idea is consistent with the
results of NMR and X-ray crystallography that yield a set of positions
$\mathbf{R}^0=\{\mathbf{r}_i^0\}$ of the $N+1$ residues to which we
refer as ``the structure'' of a protein
(NMR experiments in fact yield an ensemble of such structures). 

In GNMs a pair of residues $i,j$ within a cutoff distance (i.e. $|\mathbf{r}_i^0-
\mathbf{r}_i^0|\le r_c$) are assumed to be connected by \emph{identical} (for
sake of simplicity) but \emph{non-Hookean} springs with a constant
$K$. The interaction energy as a function of the particles'
positions $\mathbf{R}=\{\mathbf{r}_i\}$ is written as
\begin{equation}
    U_\mathrm{GNM}(\{\mathbf{r}_{ij}\})=\frac{K}{2}\sum_{\langle i,j\rangle}
    (\mathbf{r}_{ij}-\mathbf{r}_{ij}^0)^T(\mathbf{r}_{ij}-\mathbf{r}_{ij}^0),
\end{equation}
where the sum spans all connected pairs. We now introduce for
convenience the \emph{deviation from the  (equilibrium) ``structure''}, $\Delta \mathbf{R}=\{\Delta
\mathbf{r}_i\equiv\mathbf{r}_i-\mathbf{r}_i^0\}$.
The main simplifying hypothesis of the GNM is that $\Delta \mathbf{R}$
at tempertature $T$
corresponds to an isotropic Gaussian random super-vector, i.e.  
\begin{equation}
    P(\Delta \mathbf{R})=\left[(2\pi)^N
      \tilde{K}\det \Gamma^{-1}\right]^{-3/2}
    \exp\left(-\frac{\tilde{K}}{2}\Delta\mathbf{R}^T\boldsymbol{\Gamma}\Delta\mathbf{R}\right),
\end{equation}
where $\tkappa\equiv K/k_{\rm B}T$ is the dimensionless
strength (in units of thermal energy $k_{\rm B}T$) and
$\boldsymbol{\Gamma}$ is a $3(N+1)\times\nonumber 3(N+1)$ block matrix
in which each diagonal block is the connectivity (or Kirchhoff) matrix
$\Gamma$ with elements
\begin{equation}
    \Gamma_{ij}=
    \cases{
    -1, &if $i\neq j$ and  $|\mathbf{r}_i^0-\mathbf{r}_j^0|\leq r_c$\\
    0,  &if $i\neq j$ and $|\mathbf{r}_i^0-\mathbf{r}_j^0|> r_c$\\
    -\sum_{j,j\neq i}^{N+1} \Gamma_{ij}, &if i=j.\\
    }
\end{equation}
The dynamics of the beads' positions (i.e. deviations from the
equilibrium ``structure'') is assumed to follow the It\^{o} equation
\begin{equation}
    d\Delta\mathbf{R}(t)=-\xi K\mathbf{\Gamma}\Delta\mathbf{R}(t) dt+\sqrt{2D}\,d\mathbf{W}(t),
\end{equation}
where $D$ is the diffusion coefficient and $\xi\equiv D/k_{\rm B}T$
the mobility both assumed to be equal for all
beads, and  $d\mathbf{W}(t)$ is the increment of the multi-dimensional
Wiener process (i.e. Gaussian white noise)  with
zero mean and covariance $\langle dW_i(t)dW_j(t')\rangle=\delta_{ij}\delta(t-t')$.
A discussion of ANM would require a different matrix
$\mathbf{\Gamma}$ to take into account for anisotropic interactions
between beads. Moreover, the potential energy $U_{\emph ENM}$ would
depend only on distances between the residues ~\cite{bahar_normal_2010}. We do not treat this model here.

Henceforth we measure energy in units of thermal energy $k_{\rm B}T$
(i.e. $U\to U/k_{\rm B}T$),
distances in units of the cutoff distance $r_c$ (i.e. $\Delta R_i\to \Delta R_i/r_c$) and time in units of
the \emph{diffusion time}, $t_D\equiv r_c^2/D$-- the time required
for a bead with a diffusion coefficient $D$ to
diffuse a distance $r_c$ (i.e. $t\to t/t_D$).

It is convenient to pass to normal super-coordinates
$\mathbf{Q}=\{\mathbf{q}_k\}$ that diagonalize $\boldsymbol{\Gamma}$,
i.e. $\mathbf{Q}^T\boldsymbol{\Gamma}\mathbf{Q}={\rm
  diag}(\boldsymbol{\mu})$ with $(\mathbf{Q})_{ij}\equiv
Q_{ij}\mathbbm{1}$, $\mathbbm{1}$ being the $3\times 3$
identity matrix and
where the matrix $Q$ diagonalizes the Kirchoff
matrix, i.e. $Q^T\Gamma Q={\rm
  diag}(\mu_i)$, and therefore ${\rm
  diag}(\boldsymbol{\mu})_{ii}=\mu_i\mathbbm{1}$. For convenience we
let $k\in\{0,\cdots,N\}$ with $\mu_0=0$ and $Q_{i,0}$ referring to the center of
mass motion, while $i\in\{1,\cdots,N+1\}$ such that 
\begin{equation}
    \Delta\mathbf{r}_i=\sum_{k=0}^N Q_{ik}
    \mathbf{q}_k,\,\forall i\{1,\cdots,N+1\}.
    \label{distance as normal}
\end{equation}
In this notation the It\^{o} equation corresponds to the Fokker-Planck
equation describing $N$ independent isotropic three-dimensional
Ornstein–Uhlenbeck processes. Neglecting the center of mass motion we
obtain the following equation for the Green's function
(i.e. transition probability density function)
\begin{equation}
\partial_t G(\mathbf{Q}, t | \mathbf{Q}_0)=\sum_{k=1}^N \left[\partial_{\mathbf{q}_k}^2+\mu_k \partial_{\mathbf{q}_k}\mathbf{q}_k\right]G(\mathbf{Q}, t | \mathbf{Q}_0),
    \label{Markovian green function}
\end{equation}
with localized initial condition $G(\mathbf{Q}, t=0 | \mathbf{Q}_0)=\delta(\mathbf{Q}-\mathbf{Q}_0)$ 
and natural boundary conditions $\lim_{|\mathbf{Q}|\to\infty}G(\mathbf{Q}, t | \mathbf{Q}_0)=0$.
We solve Eq.~(\ref{Markovian green function}) by means of an eigendecomposition
~\cite{wilemski_diffusioncontrolled_1974} yielding
\begin{equation}
    G(\mathbf{Q}, t | \mathbf{Q}_0)=\sum_{\mathbf{N}} \Psi_{\mathbf{N}}^R(\mathbf{Q})\Psi_{\mathbf{N}}^L(\mathbf{Q_0})\ee{-\Lambda_\mathbf{N}t},
    \label{Green Markovian series}
\end{equation}
where $\Lambda_\mathbf{N}$ denote eigenvalues, $\mathbf{N}$
being a multiset of integer-triples
$\{\mathbf{n}_1,\cdots,\mathbf{n}_N\}$ with
$\mathbf{n}_i=\{n_{ix},n_{iy},n_{iz}\}$ such that
\begin{equation}
    \Lambda_\mathbf{N}=\sum_{i=1}^N (n_{ix}+n_{iy}+n_{iz})\mu_i,
    \label{many body eigenvalues}
\end{equation}
and
$\Psi_{\mathbf{N}}^L(\mathbf{Q})$ and $\Psi_{\mathbf{N}}^R(\mathbf{Q})$ are the corresponding
left and right eigenfunction given by
\begin{equation}
    \Psi_{\mathbf{N}}^L(\mathbf{Q})=\prod_{i=1}^N
  \psi_{\mathbf{n}_i}(\mathbf{q}_i), \quad \Psi_{\mathbf{N}}^R(\mathbf{Q})= P_\mathrm{eq}(\mathbf{Q})\prod_{i=1}^N \psi_{\mathbf{n}_i}(\mathbf{q}_i)
    \label{many body eigenfunctions}
\end{equation}
where
$P_\mathrm{eq}(\mathbf{Q})\prod_{i=1}^N(\mu_i/2\pi)^{3/2}\ee{-\mu_i\mathbf{q}_i^2/2}$
is the equilibrium probability density function of normal coordinates and
\begin{equation}
    \psi_{\mathbf{n}_i}(\mathbf{q}_i)=\frac{H_{n_{ix}}(\mu_i q_i^x/2)H_{n_{iy}}(\mu_i q^y_i/2)H_{n_{iz}}(\mu_i q^z_i/2)}{\sqrt{2^{n_{ix}+n_{iy}+n_{iz}}n_{ix}!n_{iy}!n_{iz}!}},
\end{equation}
where $q^x,q^y$ and $q^z$ are the components of the vector $\mathbf{q}$, and $H_n(x)$ denotes the $n$th "physicist's" Hermite polynomial~\cite{abramowitz_handbook_2013}.
Using Mehler's formula \cite{Mehler}
\begin{equation}
\sum_{n=0}^\infty\frac{(y/2)^n}{n!}H_n(x)H_n(z)=\frac{1}{\sqrt{1-y^2}}\exp\left(-\frac{y^2[x-z]^2}{1-y^2}\right)
  \label{Mehler}
\end{equation}
and recalling that $\mu_0=0$ we can also write Eq.~(\ref{Green Markovian series}) in a closed form~\cite{risken_fokker-planck_1996}
\begin{equation}
    G(\mathbf{Q},t|\mathbf{Q}_0)=\prod_{i=1}^N \left(\frac{\mu_i}{2\pi(1-\ee{-2\mu_i t})}\right)^{3/2}\exp\left(-\frac{\mu_i(\mathbf{q}_i-\mathbf{q}_{i0}\ee{-\mu_i t})^2}{2(1-\ee{-2\mu_i t})}\right),
    \label{Markovian Green closed}
\end{equation}
where the equilibrium probability density function corresponds
\begin{equation}
P_{\rm eq}(\mathbf{Q})\equiv \lim_{t\to\infty}G(\mathbf{Q},t|\mathbf{Q}_0)
  \label{eqpdf}
  \end{equation}
In what follows we will use both forms of the Green's function,
i.e. Eq.~(\ref{Green Markovian series}) and Eq.~(\ref{Markovian Green closed}).

\section{Conformational dynamics}
Throughout we are interested in conformational motions encoded in the
dynamics of some internal distance $d$, e.g. the distance between two
beads $i$ and $j$, $l=|\mathbf{r}_i-\mathbf{r}_j|$ or the distance
between the center of masses of two sets of beads $\Omega_1,\Omega_2$
with $\Omega_1\cap\Omega_2=\{0\}$,
$l_{\Omega_1,\Omega_2}=|\sum_{i\in\Omega_1}\mathbf{r}_i/{\rm card}(\Omega_1)-\sum_{i\in\Omega_2}\mathbf{r}_j/{\rm card}(\Omega_2)|$
where ${\rm card}(\Omega_i)$ is the cardinality the set
$\Omega_i$. 
Without loss of generality we may thus focus on the
distance between two arbitrary beads. Note that in absence of any
dynamics in an equilibrium at $T=0$ such a distance is constant and
equal to $d_0$. Expressed in normal coordinates we in turn have 
\begin{equation}
    \mathbf{l}\equiv\mathbf{r}_i-\mathbf{r}_j=\sum_{k=1}^N (Q_{ik}-Q_{jk})\mathbf{q}_k +\mathbf{r}_i^0-\mathbf{r}_j^0\equiv\sum_{k=1}^N A_k\mathbf{q}_k+\mathbf{d}_0,
\end{equation}
where in the second equality we have defined $A_k$ and
$\mathbf{d}_0$ and omitted the labels $i,j$ to simplify the
notation. Note, moreover, that $l\equiv |\mathbf{l}|$ and the
generalization to $l_{\Omega_1,\Omega_2}$ follows by linear superposition.

We will focus on four types of observables. The first one is the (non-Markovian)
conditional probability density of the time series of the coordinate,
$l_t$, defined as
\begin{equation}
\mathcal{G}_{d_0}(l,t|l_0)\equiv \mathbb{P} (l_t\in l{\rm
  d}l|l_{t=0}\in l_0{\rm
  d}l)=\frac{\langle\delta(l(\mathbf{Q}_t)-l)\delta(l(\mathbf{Q}_{0})-l_0)\rangle_{\mathbf{Q}_t}}{\langle\delta(l(\mathbf{Q}_{0})-l_0)\rangle_{\rm
eq}},
\label{cdf}  
\end{equation}
with $\lim_{t\to\infty}\mathcal{G}_{d_0}(l,t|l_0)\equiv\mathcal{P}_{d_0}^{\rm
  eq}(l)=\langle\delta(l(\mathbf{Q})-l)\rangle_{\rm
eq}$, 
and
where in the second equality we have used the law of conditional
probability and introduced the expectation over all Markovian paths of
the full system evolving from equilibrium $ \langle \cdot \rangle_{\mathbf{Q}_t}$, i.e. 
\begin{equation}
  \langle \mathcal{B}\rangle_{\mathbf{Q}_t}\equiv \int d\mathbf{Q}\int
  d\mathbf{Q}_0  \mathcal{B}(\mathbf{Q},\mathbf{Q}_0)G(\mathbf{Q}, t | \mathbf{Q}_0)P_{\rm eq}(\mathbf{Q}_0).
  \label{brackets1}
\end{equation}
and the expectation of any observable $\mathcal{B}(\mathbf{Q})$ over
the equilibrium measure $\langle \cdot\rangle_{\rm eq}$ is  $\langle \mathcal{B}\rangle_{\rm eq}\equiv \int d\mathbf{Q} \mathcal{B}(\mathbf{Q})P_{\rm eq}(\mathbf{Q})$.
The second observable is the normalized equilibrium autocorrelation function
\begin{equation}
\mathcal{C}_{d_0}(t)\equiv\frac{\langle l(t)l(0)\rangle-\langle
  l(t)\rangle\langle l(0)\rangle}{\langle l^2\rangle_{\rm eq}-\langle l\rangle_{\rm eq}^2}
\label{acf}  
\end{equation}
where we have introduced th expectations 
\begin{eqnarray}
  &&\langle l(t)l(0)\rangle\equiv \langle l(\mathbf{Q}_{t})l(\mathbf{Q}_{0})\rangle_{\mathbf{Q}_t}=\int_0^\infty dl\int_0^\infty
  dl_0ll_0\mathcal{G}_{d_0}(l,t|l_0)\mathcal{P}_{d_0}^{\rm eq}(l_0)\nonumber\\
 &&\langle
  l(t)\rangle\equiv \frac{\langle
    l(\mathbf{Q}_{t})\delta(l(\mathbf{Q}_{0})-l_0)\rangle_{\mathbf{Q}_t}}{\langle
    \delta(l(\mathbf{Q}_{0})-l_0)\rangle_{\rm eq}}= \int_0^\infty dl
  l\mathcal{G}_{d_0}(l,t|l_0)\nonumber\\
  &&\langle l^n\rangle_{\rm eq}\equiv  \langle l^n(\mathbf{Q})\rangle_{\rm
    eq}=\int_0^\infty dl l^n\mathcal{P}_{d_0}^{\rm eq}(l)
  \label{brackets2}
\end{eqnarray}
The third observable is the $3(N+1)\times\nonumber 3(N+1)$
position-covariance matrix~\cite{amadei_essential_1993} whose elements
are defined as
\begin{equation}
    C^{ij}_{\alpha\beta}(t,t_0)=\langle (\mathbf{r}_{i,\alpha}(t+t_0)-\mathbf{r}_{i,\alpha}^0)(\mathbf{r}_{j,\beta}(t_0)-\mathbf{r}_{j,\beta}^0) \rangle_{\mathbf{Q}_t},
\end{equation}
where $\mathbf{r}_{i,\alpha}$ is the $\alpha=\{x,y,z\}$ component of the
position vector of bead $i$, $\mathbf{r}_i$.

The fourth, time-average observable is a functional of the projected path
$l_\tau$ evolving from $l_{\tau=0}$  
called the \emph{fraction of occupation time} or ``empirical density'' \cite{lapolla_spectral_2020}
\begin{equation}
\theta_{d_0}(l;t)\equiv t^{-1}\int_0^t\delta(l_\tau-l)d\tau.
\label{occt}  
\end{equation}
Note that all observables defined above are assumed to evolve from
equilibrium. However, except for  $C^{ij}_{\alpha\beta}(t,t_0)$, the initial
distribution in fact corresponds to equilibrium  constrained to
a given value of the tagged distance $l_0$, i.e. from \emph{all} those
equilibrium configurations drawn from $P_{\rm eq}(\mathbf{Q})$ that
are compatible with $l_0$. This introduces memory in the dynamics of
$l_t$ \cite{lapolla_manifestations_2019}.

\subsection{Projected propagator}
The non-Markovian projected propagator $\mathcal{G}_{d_0}(l,t|l_0)$ defined in Eq.~(\ref{cdf}) denotes the probability
density that the distance between the  two tagged beads is equal to
$l$ at time $t$ given that it was initially equal to
$l_0$. Introducing the auxiliary functions
\begin{equation}
\eta_t\equiv \sum_{k=1}^N\frac{A_k^2}{2\mu_k}\ee{-\mu_kt},\quad
\Xi_{t}(d_0,l,l')
\equiv\erfi\left(\frac{d_0(\eta_0-\eta_t)+\eta_t(l+l')}{2\sqrt{\eta_t(\eta_0^2-\eta_t^2)}}\right)
\label{aux}  
\end{equation}
we find (for details of the calculation see  \ref{appendix:pdf})
\begin{eqnarray}
\mathcal{P}_{d_0}^{\rm
  eq}(l_0)
\mathcal{G}_{d_0}(l,t|l_0)=&&\frac{ll_0\exp\left(-\frac{(l^2+l_0^2)\eta_t+(\eta_0-\eta_t)d_0^2}{4\eta_t(\eta_0-\eta_t)}\right)}{8\sqrt{\pi
    \eta_t}d_0(\eta_0-\eta_t)}[\Xi_{t}(d_0,-l,-l_0)-\nonumber\\
  &&\Xi_{t}(d_0,-l,l_0)+\Xi_{t}(d_0,l,l_0)+\Xi_{t}(-d_0,-l,l_0)]
\label{joint closed}
\end{eqnarray}
where $\erfi(x)$ is the imaginary error function
\cite{abramowitz_handbook_2013} and
\begin{equation}
 \mathcal{P}^{\rm eq}_{d_0}(l)=\frac{l}{d_0}\frac{{\rm e}^{-(l^2+d_0^2)/4\eta_0}}{\sqrt{\pi\eta_0}}\sinh\left(\frac{ld_0}{2\eta_0}\right),
 \label{non markovian pdf}
\end{equation}
We also derive the spectral expansion of $\mathcal{G}_{d_0}(l,t|l_0)$
that reads  (see \ref{appendix:green series}) 
\begin{equation}
    \mathcal{G}_{d_0}(l,t|l_0)=V_{\mathbf{00}}(l_0;d_0)^{-1}\sum_{\mathbf{N}}V_{\mathbf{0N}}(l;d_0)V_{\mathbf{N0}}(l_0;d_0)\ee{-\Lambda_Nt},
    \label{green series}
\end{equation}
where the overlap elements $V_{\mathbf{0N}}$ and $V_{\mathbf{N0}}$ admit a
closed-form expression that is, however, somewhat complicated and thus
given in \ref{appendix:green series}. Note that ``the ground state''
element is simple and corresponds to
$V_{\mathbf{00}}(l;d_0)=\mathcal{P}^{\rm eq}_{d_0}(l)$.

\subsection{Equilibrium distance autocorrelation function}
The (normalized) autocorrelation function defined in Eq.~(\ref{acf})
is made explicit by means of the following results
\begin{eqnarray}
   &\langle l\rangle_{\rm eq}=  2\sqrt{\frac{\eta_0}{\pi}} \ee{-d_0^2/4\eta_0}+\left(d_0+\frac{2\eta_0}{d_0}\right)\mathrm{erf}\left(\frac{d_0}{2\sqrt{\eta_0}}\right),\\
  &\langle l^2\rangle_{\rm eq}=d_0^2+6\eta_0
  \label{explicit}
\end{eqnarray}
where ${\rm erf}$ is the error function. Eqs.~(\ref{explicit}) follow
from direct integration of the last line of Eq.~(\ref{brackets2})
with the aid of Eq.~(\ref{non markovian pdf}). Conversely,  an analytic computation of
$\langle l(t) l(0)\rangle$ is possible only using the spectral
expansion in Eq.~(\ref{green series}) and yields, 
\begin{eqnarray}
    &\langle l(t) l(0)\rangle=\sum_{\mathbf{N}}\mathcal{V}^{d_0}_{\mathbf{0N}}\mathcal{V}^{d_0}_{\mathbf{N0}}\ee{-\Lambda_Nt}, \label{autocorr series}\\
    &\mathcal{V}^{d_0}_{\mathbf{0N}}=\int_0^\infty dl l
  V_{\mathbf{0N}}(l;d_0),\quad 
  \mathcal{V}^{d_0}_{\mathbf{N0}}=\int_0^\infty dll  V_{\mathbf{N0}}(l;d_0).
  \label{elem2}
\end{eqnarray}
The analytic expression of the coefficients $\mathcal{V}^{d_0}_{\mathbf{0N}}$ is lengthy and can be
found in \ref{appendix: autocorrelation}. Plugging Eqs.~(\ref{elem2})
and Eq.~(\ref{explicit}) into
Eq.~(\ref{acf}) delivers an exact
analytical result for the equilibrium distance autocorrelation
function $\mathcal{C}_{d_0}(t)$. Alternatively one may also evaluate
$\mathcal{C}_{d_0}(t)$ by numerical integration of the first line of
Eq.~(\ref{brackets1}) using Eq.~(\ref{joint closed}), which may in
fact be numerically more convenient than implementing the analytical solution.

\subsection{Position covariance matrix}
In the analysis of atomistic Molecular Dynamics (MD) simulations one often
focuses on the position covariance matrix $C^{ij}_{\alpha
  \beta}(t,t_0)$ ~\cite{amadei_essential_1993} and its
eigendecomposition. The trajectory derived from an MD stimulation is
then projected on the eigenvector (or principal component)
corresponding the largest eigenvalue of the covariance matrix with the
aim to identify the most important (potentially functionally relevant)
motion in a protein~\cite{amadei_essential_1993}. To facilitate a
comparison between the aforementioned analysis of MD simulation with
GNM we compute $C^{ij}_{\alpha
  \beta}(t,t_0)$ analytically. Passing as before to normal
coordinates we find
\begin{equation}
    C^{ij}_{\alpha \beta}(t,t_0)=\langle \sum_{k=1}^NQ_{ik} q_{k\alpha}(t+t_0) \sum_{l=1}^NQ_{jl} q_{l\beta}(t_0) \rangle,
    \label{covariance element}
\end{equation}
where the matrix elements $Q_{ij}$ do not depend on the spatial
coordinate because the GNM is isotropic. Each process
$\mathbf{q}_{k,\alpha}$ corresponds to an independent
Ornstein–Uhlenbeck process, i.e. the solution of the It\^{o}
integral~\cite{gardiner_c.w._handbook_1985} (setting all constant to unity)
\begin{equation}
    q_{k\alpha}(t)=\sqrt{2}\int_{0}^t \mathrm{e}^{-\mu_k(t-s)}dW_{k\alpha}(s).
\end{equation}
Since by construction (i.e. as a result of isotropy) only the elements of the same spatial coordinate
for any given normal mode survive the averaging in Eq.~(\ref{covariance element}), the elements of the covariance
matrix read explicitly
\begin{equation}
    C^{ij}_{\alpha \alpha}(t,t_0)=\sum_{k=1}^N
    \frac{Q_{ik}Q_{jk}}{\mu_k}\mathrm{e}^{-\mu_k|t-t_0|}.
    \label{eq:covmat}
\end{equation}
Obviously $C^{ij}_{\alpha \alpha}(t,t_0)$  is stationary (i.e. depends
only on the time difference, $C^{ij}_{\alpha \alpha}(t,t_0)=C^{ij}_{\alpha \alpha}(|t-t_0|)$).

\subsection{Fluctuations of occupation time}
Single molecule experiments typically probe time-averaged observables.
For example, F\"{o}rster resonance energy transfer (FRET)
~\cite{truong_use_2001} and plasmon ruler
experiments~\cite{ye_conformational_2018} have been used to extract
information about conformational motions of macro-molecules. A
fundamental quantity to that underlies this kind of observables is the
fraction of occupation time, $\theta_{d_0}(l;t)$, defined in Eq.~(\ref{occt})
\cite{kac_m_distributions_1949,yen_local_2013,majumdar_local_2002,
  wadia_brownian_2006,lapolla_unfolding_2018,lapolla_spectral_2020} --
the random fraction of time a time-series (in our case an internal
distance between two beads or between two center of masses) of length
$t$ attains a given value of $l$.

In previous publications we have shown how to obtain the mean and the
variance of
$\theta_{d_0}(l;t)$~\cite{lapolla_unfolding_2018,lapolla_spectral_2020}.
Along these lines we here focus on the mean, $\langle
\theta_{d_0}(l;t)\rangle$,  and the variance, $\sigma^2_{\theta;
  d_0}(l;t)\equiv \langle
\theta_{d_0}^2(l;t)\rangle-\langle
\theta_{d_0}(l;t)\rangle^2$, of the
occupation time fraction at equilibrium that read, respectively (for a derivation see \ref{appendix:varianceLT
  calculation})
\begin{eqnarray}
    &\langle\theta_{d_0}(l)\rangle=\mathcal{P}^{\rm eq}_{d_0}(l),\\
    &\sigma_{\theta; d_0}^2(l,t)=\frac{2}{t}\sum_{\mathbf{N}\neq\mathbf{0}}\frac{V_{\mathbf{0N}}(d;d_0)V_\mathbf{N0}(d;d_0)}{\Lambda_\mathbf{N}}\left(1-\frac{1-\ee{-\Lambda_\mathbf{N}t}}{\Lambda_\mathbf{N}t}\right).
    \label{local-time variance}
\end{eqnarray}
Note that $\langle
\theta_{d_0}(l;t)\rangle$  corresponds to the equilibrium probability
density for all times $t$ since we are considering an ergodic system
evolving from equilibrium initial conditions. The variance of the
occupation time fraction can equivalently be obtained from (see e.g.~\cite{gopich_single-macromolecule_2003})
\begin{equation}
    \sigma_{\theta; d_0}^2(l,t)=\frac{2}{t}\mathcal{P}^{\rm eq}_{d_0}(l)\left[\int_0^t (1-\tau/t) \mathcal{G}_{d_0}(l,\tau|l)-\mathcal{P}^{\rm eq}_{d_0}(l)\right]d\tau.
    \label{variance explict integral}
\end{equation}
The integral in Eq.~(\ref{variance explict integral}) does not admit
an explicit solution. However, it can easily be computed via numerical
quadrature. Moreover, it is possible to expand
$\mathcal{G}_{d_0}(l,\tau|l)$ for short times (details are given
in~\ref{appendix:short time}) yielding the \emph{small deviation limit}
\begin{equation}
  \mathcal{G}_{d_0}(l,t|l)\stackrel{t\to
    0}{=}2\sqrt{\frac{1}{\pi}}\left(\frac{2}{\sqrt{\kappa
      t}}+\frac{\sqrt{\kappa t}}{l^2}\right)+\mathcal{O}(t^{3/2})
\label{small}  
\end{equation}
where we have introduced the shorthand notation $\kappa=\sum_{k=1}^N A_k^2$.
Plugging Eq.~(\ref{small}) into Eq.~(\ref{variance explict integral})
and performing the integral in turn yields
\begin{equation}
    \sigma_{\theta;d_0}^2(d,t)\stackrel{t\to0}{\simeq}2\mathcal{P}^{\rm eq}_{d_0}(l)\left(\frac{8}{3\sqrt{\kappa\pi
    t}}+\frac{4}{15l^2}\sqrt{\frac{\kappa
        t}{\pi}}-\mathcal{P}^{\rm eq}_{d_0}(l)\right).
\label{SDEV}    
\end{equation}
Since the dynamics of every stable system at equilibrium can be
``linearized'' for sufficiently small times $t$ the \emph{small
deviation} asymptotic in Eqs.~(\ref{SDEV}) and (\ref{small}) is in fact a general result for the
 (large) fluctuations of $\theta_{d_0}(l;t)$ at sufficiently short times.

\section{Examples}
We now apply the result of the previous section to the analysis of a Gaussian
Network Model of a protein called adenylate kinase and the analysis of
toy-model mechanical frames.

\subsection{Gaussian Network Model of adenylate kinase}
\begin{figure}
    \centering
    \includegraphics[width=0.98\textwidth]{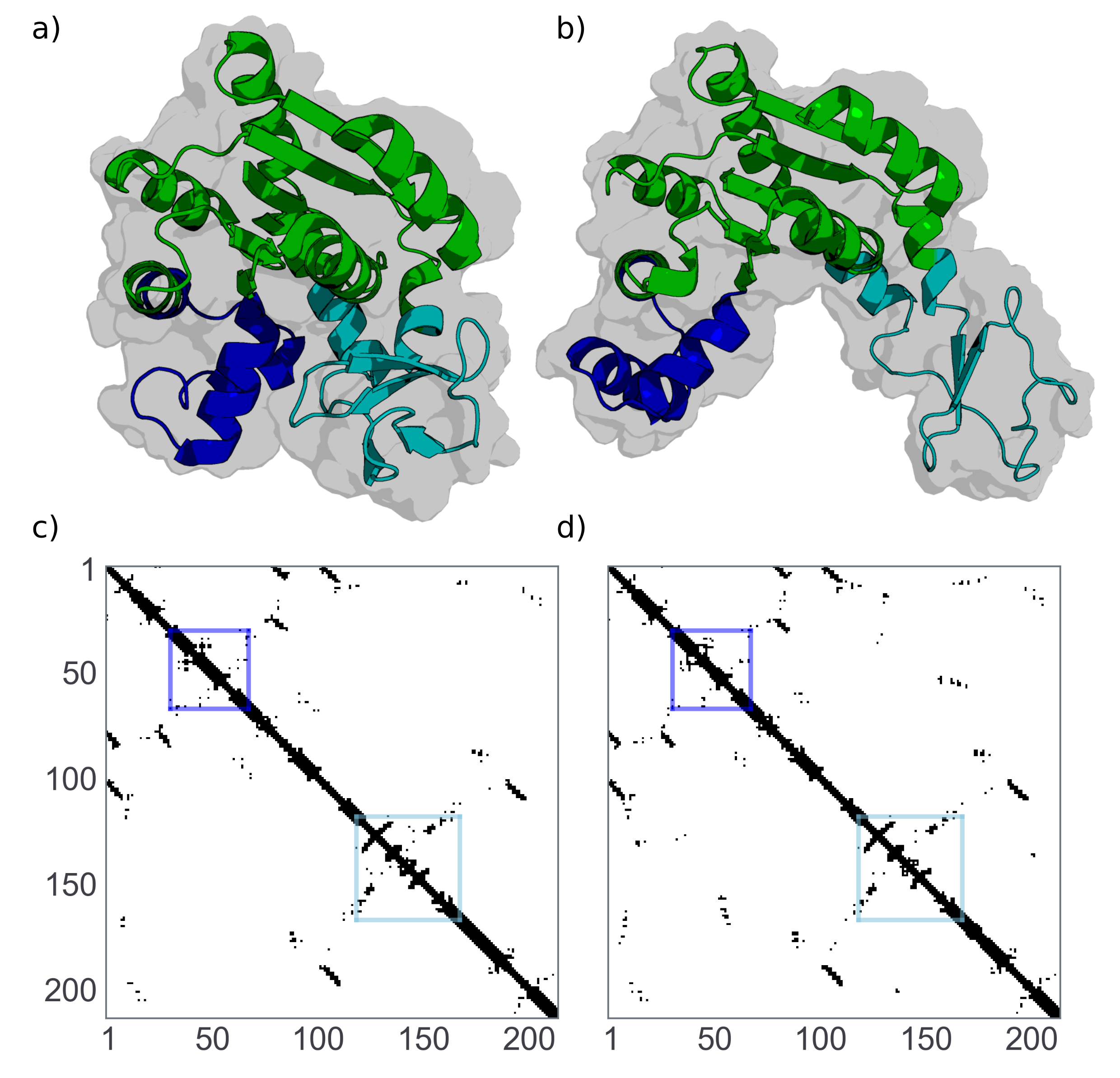}
    \caption{Panels (a) and (b) depict a cartoon and the molecular
      surface (gray) of the two protein structures, called (a) ``the closed'' configuration 1AKE 
    and (b) ``the open'' configuration 4AKE. 
    Panels (c) and (d) show the corresponding connectivity matrices
    for 1AKE and 4AKE, respectively. The blue and cyan square enclose,
    respectively, the NMP and LID residues. The cutoff distance used to obtain these matrices was $8$\,\AA.}
    \label{fig:structure2GNM}
\end{figure}

Adenylate kinase (ADK) is an enzyme catalyzing the reversible
phosphorylation reaction that transforms adenosine monophosphate (AMP)
to adenosine triphosphate (ATP). The structure of ADK has been
resolved using X-ray crystallography that uncovered two distinct
conformations of the protein that are deposited in the Protein Data
Bank (PDB ID: 1AKE~\cite{muller_structure_1992} and PDB ID:
4AKE~\cite{muller_adenylate_1996}) and shown in Fig.~\ref{fig:structure2GNM}.

ADK consists of  $214$ residues divided in 3 macro-domains called CORE
(residues $1-29$, $68-116$, and $160-214$), LID (residues $118-160$),
and NMP (residues $30-67$). Distinct studies suggest the function to
be coupled to open-closed transitions of both, LID and NMP domains
with respect to the CORE domain~\cite{muller_structure_1992,
  muller_adenylate_1996}. These transitions have been observed even in
absence of nucleotides~\cite{henzler-wildman_intrinsic_2007,
  hanson_illuminating_2007}. However, there is a lively debate in the
biophysical community about the precise mechanism and rate-limiting
steps in the catalytic function of ADK ~\cite{zheng_multiple_2018}.

Here we analyze the autocorrelation functions of distances between the
center of mass of LID, NMP, and CORE
using the results described in the previous sections. Note that
each GNM describes only a single stable structure and therefore cannot
capture transitions between the two structures. Nevertheless, the
comparison between the two respective GNMs may highlight some
differences of the dynamics around the two distinct stable minima.

We obtain the connectivity matrices (shown in
Fig.~\ref{fig:structure2GNM}) of the two GNMs using the Prody
package~\cite{bakan_prody_2011} with a cutoff distance
$r_c=8$\,\AA. The static (zero-temperature) distances between the center of masses
of the three domains in both structures are given in Table~\ref{tab:distances}.
\begin{table}[]
  \caption{Distance between the center of masses of the three domains for both structures of ADK. All distances are expressed in units of the cutoff distance $r_c=8$\,\AA.}
    \centering
    \begin{tabular}{c|c|c}
          $d_0[r_c]$ & 1AKE & 4AKE\\
          \hline
          CORE-LID & $2.6$ & $3.8$\\
          CORE-NMP & $2.3$ & $2.7$\\
          LID-NMP & $2.6$ & $4.5$
    \end{tabular}
    \label{tab:distances}
\end{table}

\begin{figure}
    \centering
    \includegraphics[width=0.98\textwidth]{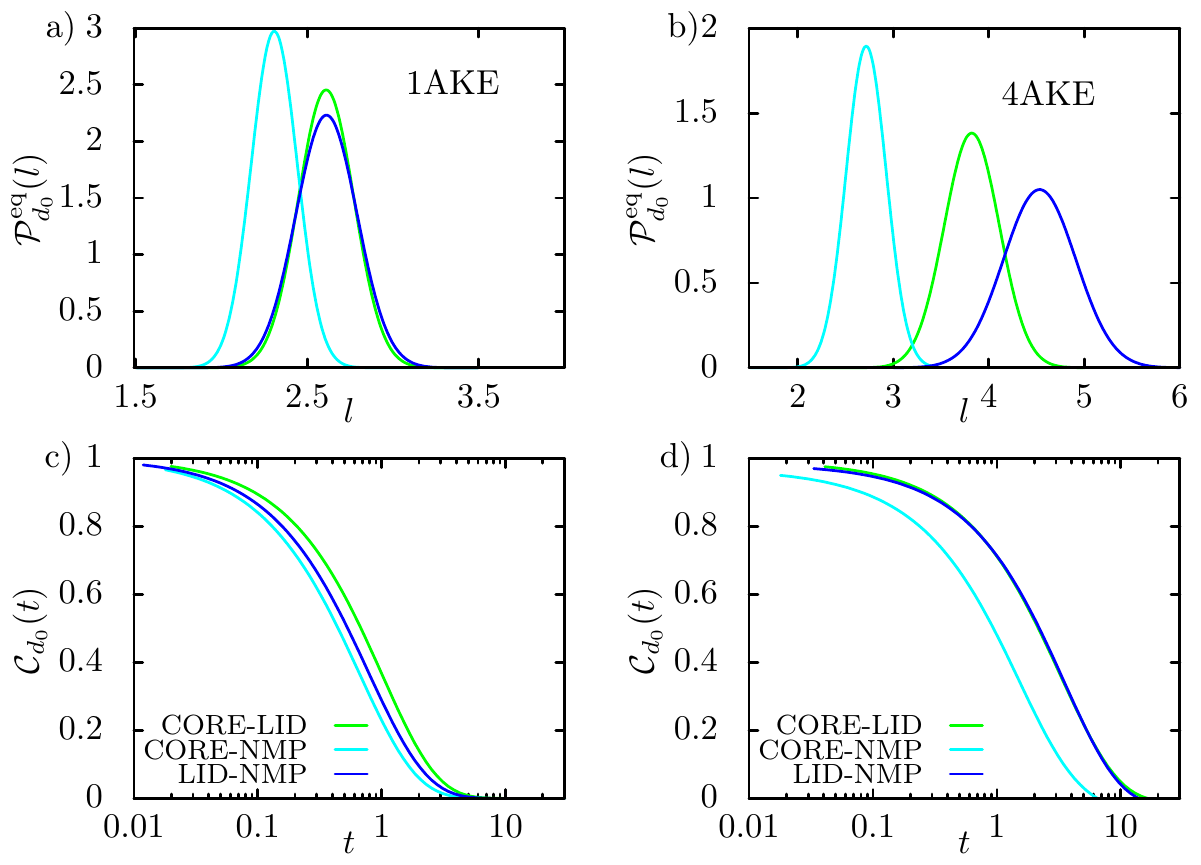}
    \caption{Panels (a) and (b) show the equilibrium probability
      density function for the three center-of-mass distances for both
      structures (see Table~\ref{tab:distances} for the numerical
      values of  $d_0$). Panels (c) and (d) depict the respective
      distance autocorrelation functions. Note that the first
      structure is more compact and its $\mathcal{C}_{d_0}(t)$
      decorrelates faster. Moreover, panel (d) reveals that the
      CORE-NMP distance decorrelates faster than the remaining two distances.}
    \label{fig:ake}
\end{figure}

Fig.~\ref{fig:ake} shows the equilibrium probability density function
$\mathcal{P}^{\rm eq}_{d_0}(l)$ (panels a and b) as well as the
autocorrelation function $\mathcal{C}_{d_0}(t)$  (panels c and d)  for all considered
distances of the two GNMs representing the two conformational states
of ADK.  The structure 1AKE is evidently more compact than 4AKE and
its corresponding autocorrelation functions consistently decay
faster. Moreover, the CORE-NMP distance autocorrelation function
decays faster compared to the other two distances whose
autocorrelation functions are almost identical (see Fig.~\ref{fig:ake}d).
This difference in relaxation is a result of differences in the
respective projection, i.e. whereas the eigenvalues of the underlying
generator are identical (see Eq.~(\ref{elem2})) the numerical
coefficients $\mathcal{V}^{d_0}_{\mathbf{0N}}$ and
$\mathcal{V}^{d_0}_{\mathbf{N0}}$ depend strongly on the particular
type of projection and thus modify the relaxation rate substantially~\cite{lapolla_toolbox_2021}.

The lines in Figs.~\ref{fig:ake}c) and~\ref{fig:ake}d) have been
obtained by means of a numerical integration of the first line of
Eq.~(\ref{brackets2}) using the Gauss-Kronrod
quadrature~\cite{noauthor_httpswwwboostorgdoclibs1_73_0libsmathdochtmlmath_toolkitgauss_kronrodhtml_2020}. Unfortunately
the evaluation of the integrand is challenging for very short-times
because it is a function sharply peaked along the diagonal of the
$l,l_0$-plane. This feature prohibits us to obtain reliably (that is, due to numerical imprecision) the autocorrelation function for very short times.  

\begin{figure}
    \centering
    \includegraphics[width=0.98\textwidth]{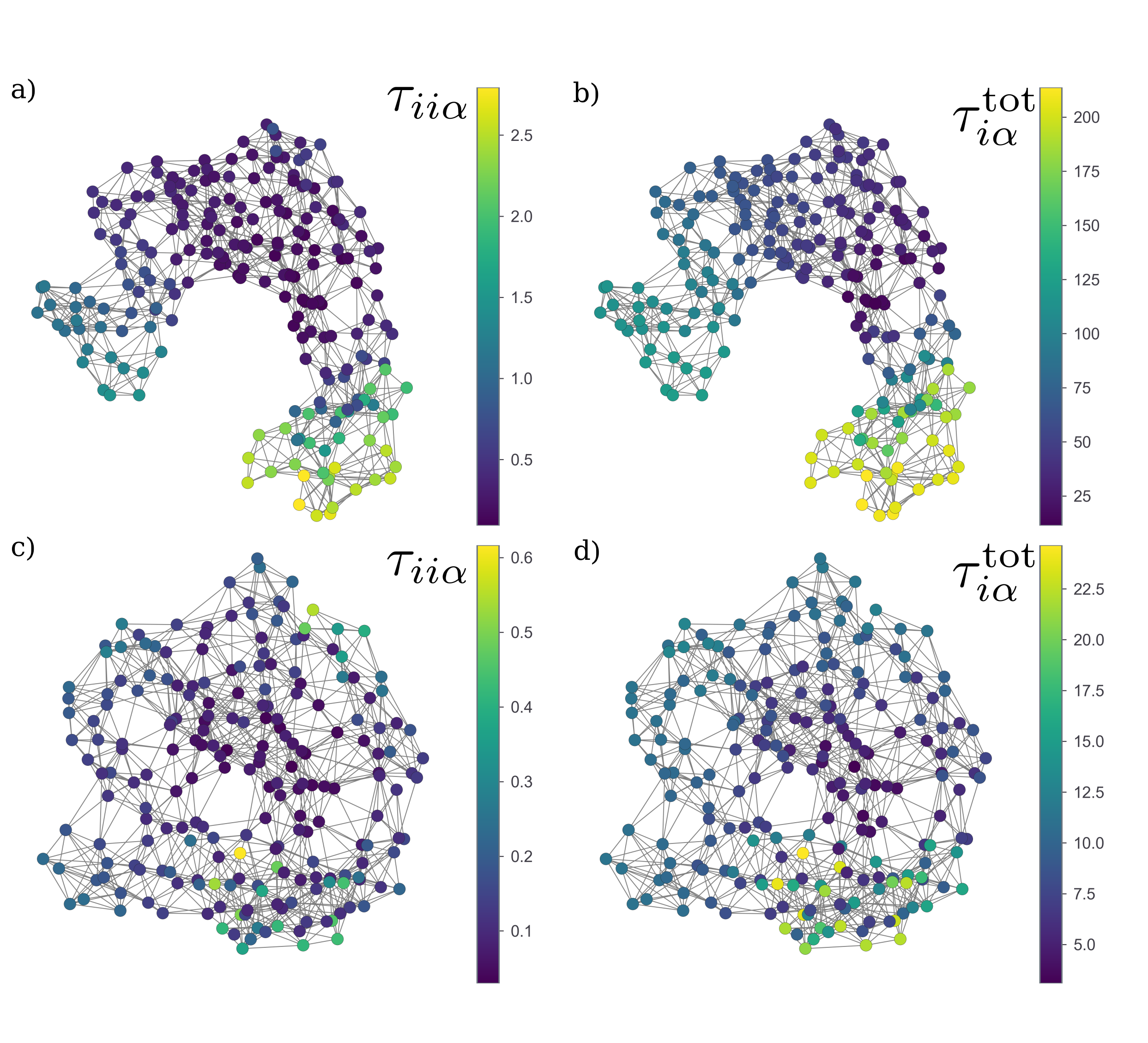}
    \caption{(a) and (c) depict $\tau_{ii\alpha}$ and (b) and (d) $\tau^{\rm
  tot}_{i,\alpha}$ for each bead in the
      4AKE and 1AKE structures, respectively. Notably,  the beads in
      the LID and NMP domains in the 4AKE structure display a
      particularly large covariance-time.}
    \label{fig:var_covar}
\end{figure}

Next we inspect the covariance matrix in Eq.~(\ref{eq:covmat}) to identify the dominant,
potentially functional important, motions in ADK. In order to reduce
the information content while retaining the most essential physics
about the extent of local fluctuations and how much the motion of each bead correlated to the motion of
other beads we introduce the \emph{covariance-time} defined as
\begin{equation}
    \tau_{ij\alpha}=\int_0^\infty C^{ij}_{\alpha\alpha}(t) dt,
\end{equation}
which may be interpreted in a manner analogous to the correlation
time~\cite{lipari_model-free_1982, perico_positional_1993,
  lapolla_single-file_2020}, i.e. as a measure of how much the motion
between the beads $i$ and $j$ is correlated over time.  To measure how
much the motion of a single bead is correlated with the rest of the
system we consider the total the total covariance-time $\tau^{\rm
  tot}_{i,\alpha}\equiv\sum_{j\ne i}
|\tau_{ij\alpha}|$. Conversely, the total variance-time is quantified
directly by $\tau_{ii\alpha}$. Note that the model is isotropic and
thus independent of $\alpha$. The results are shown in Fig.~\ref{fig:var_covar}.

Notably, one can immediately observe that those residues that are
involved in the large-scale open-closed motion (i.e. residues with a
large $\tau_{ii\alpha}$) also participate in correlated motions
denoted by large values $\tau^{\rm
  tot}_{i,\alpha}$. For the open structure 4AKE (see Fig.~\ref{fig:var_covar} a
and b) the two ends of the LID and NMP domains move in a particularly
correlated fashion. These residues are in fact those that move towards
the core region in the functional open-closed motion of the
protein~\cite{henzler-wildman_intrinsic_2007,
  hanson_illuminating_2007}. 
A remnant of this collective motion can also be seen in the closed structure 1AKE (Fig.~\ref{fig:var_covar} c
and d), where  the same beads as in 4AKE have a larger $\tau^{\rm
  tot}_{i,\alpha}$. This is likely a result of a higher local connectivity.

\subsection{Simple mechanical frames}
Although GNMs were originally developed to describe proteins they can
in fact be used to model any mechanical system in which some underlying network of links
imposes constraints on the position of nodes while allowing small,
Gaussian fluctuations driven by thermal noise. Examples may
include nano-machines such as piezoelectric actuators that move
probe-tips in atomic force microscopes \cite{tian_design_2010,
  sierra_review_2005}. 

In the generic context of ``mechanical frames'' the theory of
structural rigidity deals with the question of whether frames are
rigid or not \cite{crapo_structural_1979}. A frame is said to be rigid if
one cannot change the distance between pairs of nodes without
simultaneously altering the length of at least one connection. A
structure that is not rigid is in turn said to allow for inextensional
mechanisms. These arise 
due to a too low number or a particular arrangement of links. In
addition, in frames with redundant links there exist states of
self-stress. Under given circumstances these states of self-stress
impart stiffness to inextensional mechanisms
\cite{calladine_buckminster_1978}.

\begin{figure}
    \centering
    \includegraphics[width=0.98\textwidth]{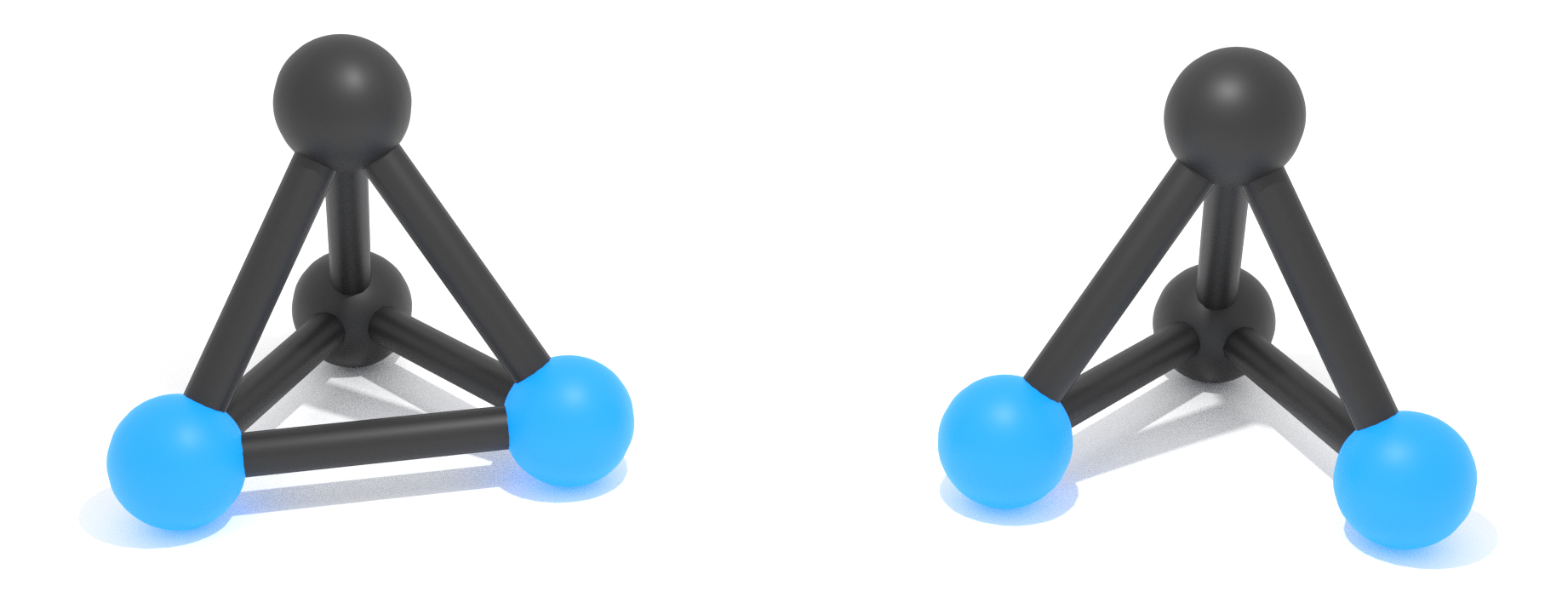}
    \caption{A schematic representation of a stable (left) and an
      unstable (right) frame we consider below.}
    \label{fig:mech_struct}
\end{figure}

As anticipated by Maxwell  such a classification of mechanical frames
is often non-trivial and may require more information than encoded in
the topology of the network \cite{maxwell_l_1864}. A complete analysis
of the mechanisms of a given frame can be obtained by a ``singular
value decomposition`` of the respective Equilibrium Matrix
$\mathbf{A}$ \cite{pellegrino_matrix_1986} that relates forces $\mathbf{f}$ on the nodes with tensions $\mathbf{t}$ in the links
\begin{eqnarray}
    \mathbf{A} \mathbf{t} = \mathbf{f}
    .
    \label{eq:eqmat}
\end{eqnarray}
Singular value decomposition of $\mathbf{A}$ allows (amongst other
things) to determine the rank $r$ of $\mathbf{A}$ and thereby the
number of inextensional mechanisms $m$ and states of self-stress $s$
via $s = b - r$ and $m = 3 j - 6 - r $, where $j$ is the number of
joints and $b$ the number of links in the structure, and note that
there are in general 6 rigid-body motions in 3 spatial dimensions.
Maxwell's well-known formula $b = 3j -6$ is then extended to:
\begin{eqnarray}
    b - 3 j + 6 = s - m
    .
    \label{eq:sm}
\end{eqnarray}
To illustrate the concept we consider two toy-model frames
depicted in Fig.~\ref{fig:mech_struct}. Both have $j=4$ nodes and
$s=0$ states of self-stress. The rigid structure with $b=6$ links has
no inextensional mechanism (i.e. $6-12-6 = 0 - 0$) while the structure
with $b=5$ links has exactly $m=1$ mechanism (i.e. $5-12-6 = 0-1$).

To highlight the r\^ole of rigidity and to investigate the effect of a heat-bath we first analyze the
autocorrelation function between the blue beads (see
Fig.~\ref{fig:mech_struct}) as a function of the rest-length
$d_0$. Notably, in a GNM such distance fluctuations do
\emph{not} depend on the equilibrium structure $\mathbf{R}^0$. Only
the  equilibrium distance between the tagged beads,
$d_0=|\mathbf{r}_i^0-\mathbf{r}_j^0|$,  is relevant. In turn there is
a redundancy --  many distinct equilibrium structures $\mathbf{R}^0$
may yield the same result that depends only on the connectivity matrix
$\boldsymbol{\Gamma}$ and $d_0$. 

\begin{figure}
    \centering
    \includegraphics[width=0.98\textwidth]{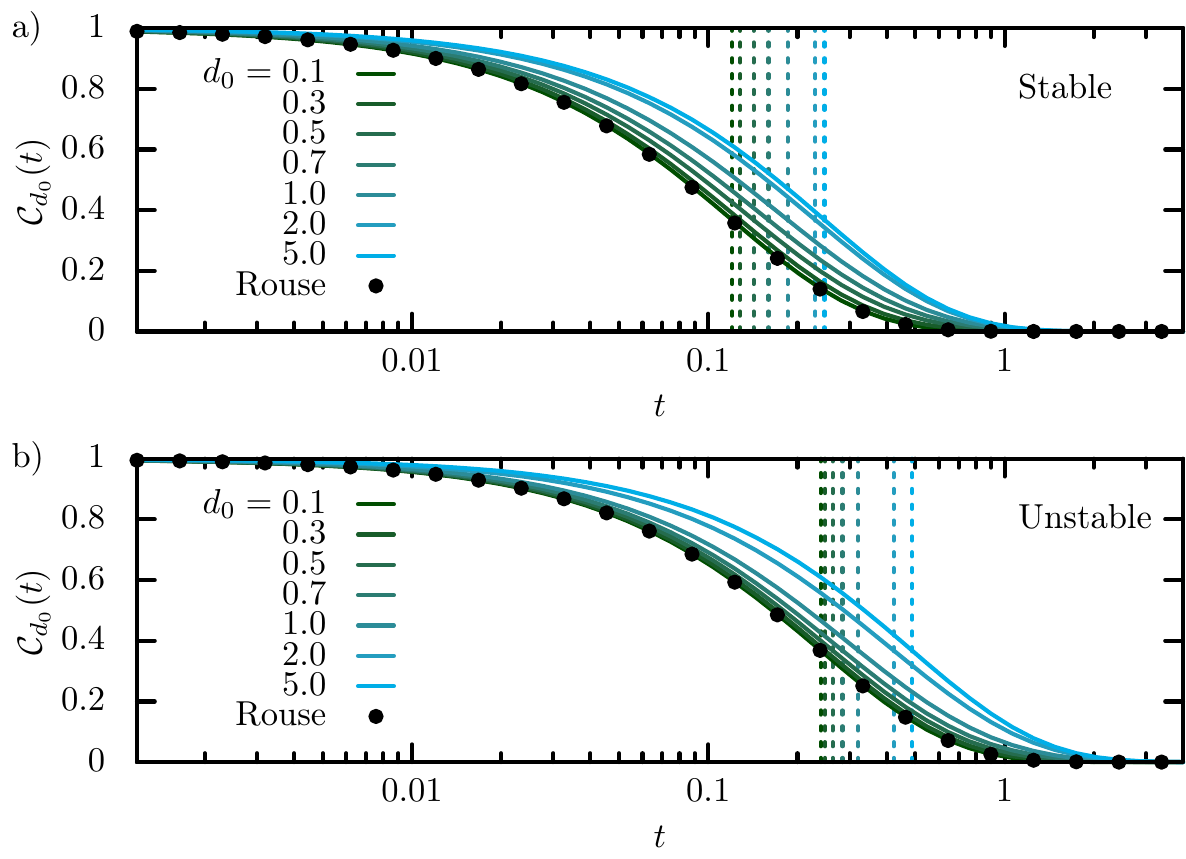}
    \caption{Distance autocorrelation function $\mathcal{C}_{d_0}(t)$
      for various values of the rest length  $d_0$ for the rigid (top
      panel) and non-rigid (bottom panel) frames depicted in
      Fig.~\ref{fig:mech_struct}. The black dots depict
      $\mathcal{C}_{d_0}(t)$ in the 
      Rouse limit $d_0=0$ (see Appendix~\ref{appendix:
        autocorrelation} for details).  The vertical dashed lines
      corresponds to the time $t_c$ at which $\mathcal{C}_{d_0}(t_c)
      =\mathrm{e}^{-1}$. Note that the unstable structure relaxes slower.}
    \label{fig:autocorr}
\end{figure}

The (normalized) distance autocorrelation function
$\mathcal{C}_{d_0}(t)$ (see Eq.~(\ref{acf})) for the two frames is
shown Fig.~\ref{fig:autocorr}. For $d_0 \lesssim 0.5$ (in
dimensionless units) $\mathcal{C}_{d_0}(t)$ depends only very weakly
on $d_0$. For larger values of $d_0$ the relaxation time (see dashed
vertical lines in Fig.~\ref{fig:autocorr}) increases. This observation
may be explained by noticing that entropy dominates the motion for
small $d_0$. That is, in the limit of small $d_0$ the rest length may
be neglected and the ``Rouse limit'' suffices to explain the dynamics
essentially quantitatively. Conversely, as $d_0$ increases a certain
``stiffness'' emerges in the frame and the (random) oscillations become localized
around the equilibrium value $d_0$.  
Note that the entropic contribution to $\mathcal{C}_{d_0}(t)$ is more
important for the non-rigid frame (see right panel in
Fig.~\ref{fig:autocorr}) as we increase the value of $d_0$ (see
Fig.~\ref{fig:autocorr}b)). Conversely, the departure from the Rouse
limit towards the ``large stiffness" case is faster in the stable frame
(see Fig.~\ref{fig:autocorr}a)). A larger $d_0$ leads to a slower decay of the autocorrelation function $\mathcal{C}_{d_0}(t)$.

\begin{figure}
    \centering
    \includegraphics{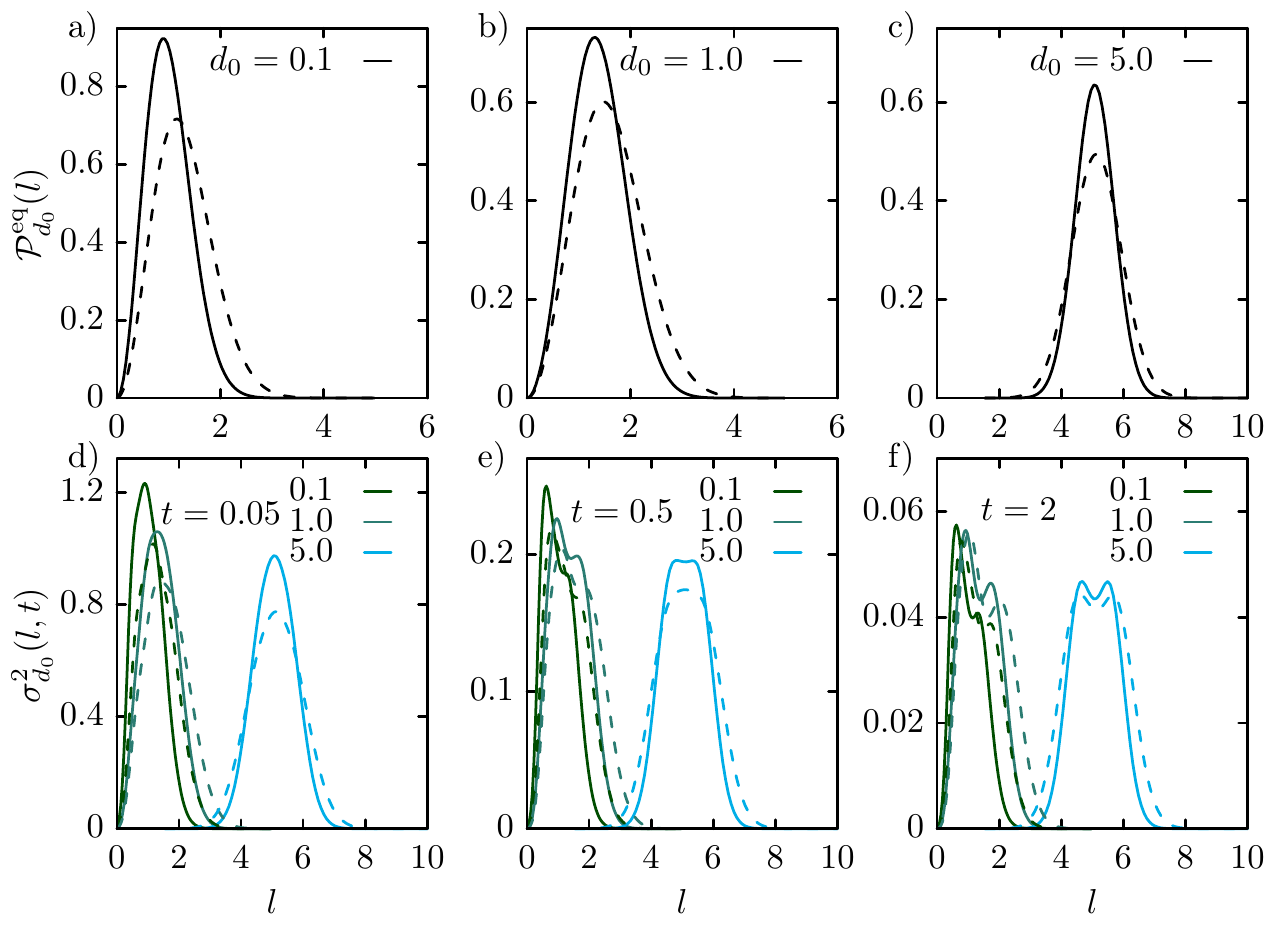}
    \caption{Panels a-c show the equilibrium probability
      density $\mathcal{P}^{\rm eq}_{d_0}(l)$ for the stable 
      (full lines) and unstable (dashed lines) structure for
      several values of $d_0$. Panels e-f depict the variance of the
      occupation time $\sigma_{\theta;
  d_0}^2(l,t)$ for the stable 
      (full lines) and unstable (dashed lines) structure,
      respectively, for different values of $d_0$.  The length of the
      trajectory $t$ increases from d to f.}
    \label{fig:LTS}
\end{figure}

Next we consider the fraction of occupation time $\theta_{d_0}(l;t)$
\cite{lapolla_spectral_2020}. We assume that the initial condition
evolves from equilibrium and therefore
$\langle\theta_{d_0}(l;t)\rangle=\mathcal{P}^{\rm eq}_{d_0}(l)$
whereas $\sigma_{\theta; d_0}^2(l,t)$ depends on time (see
Eq.~(\ref{local-time variance}) as well as
\cite{lapolla_unfolding_2018, lapolla_spectral_2020}). The
aforementioned dominance of the entropic (heat bath) contribution at
small values of $d_0$ is
also noticeable the the fluctuations of $\theta_{d_0}(l;t)$ as
depicted in Fig.~\ref{fig:LTS}. Notably, as $d_0$
increases the support of $\sigma_{\theta; d_0}^2(l,t)$ progressively
shifts towards larger $l$ and concentrates near $d_0$. 

Notably, the variance of the occupation time fraction $\sigma_{\theta;
  d_0}^2(l,t)$ changes shape from unimodal shape at short times $t$ to
bimodal at long $t$. Such a behavior is characteristic for stochastic
process in spatial confinement \cite{lapolla_spectral_2020},
i.e. fluctuations of $\theta_{d_0}(l;t)$ are larger in the vicinity of
confining boundaries (even if these boundaries are ``soft''). 

Moreover as $d_0$ increases the shape of both, $\mathcal{P}^{\rm
  eq}_{d_0}(l)$ 
as well as $\sigma_{\theta;
  d_0}^2(l,t)$ becomes more symmetric. The reason seems to be that the effect
of the confining boundary at $l=0$ becomes irrelevant as the support
of $\sigma_{\theta;
  d_0}^2(l,t)$
begins to concentrate near a substantial $d_0$.
In other words although the projection of the dynamics of a link in 3-dimensional
space onto a (1-dimensional) distance destroys the Gaussian behaviour,
the latter becomes (partially) restored at large values of $d_0$.

\section{Conclusions}
We presented analytical results (up to a numerical diagonalization of a
symmetric matrix) for a selection of relevant time- and
ensemble-average physical observables in the Gaussian Network
Model (GNM). One may think of GNM as certain generalization of the
Rouse model to networks with links with a certain degree of
extensional and rotational stiffness. We determined a set of
coarse-grained observables -- internal distances -- that may be of interest in the analysis of
GNM in the context of internal motions in proteins or mechanical frames in contact with a
heat bath. We hope that our results will enable and motivate a more
systematic analysis of GNM derived from
proteins~\cite{bakan_prody_2011}. To this end a C++ computer code is
provided in the Supplementary material that implements all result (for
more details about the implementation see Appendix~\ref{appendix:computer evaluation}).

\section*{Acknowledgments}
The authors thank David Hartich and Lars Bock for the useful
discussions. The financial support from the German Research Foundation (DFG) through the Emmy Noether Program GO 2762/1-1 to AG is gratefully acknowledged.

\appendix
\section{Derivation of the equilibrium probability density}
\label{appendix:pdf}
The equilibrium probability density function of any link-vector
$\mathbf{l}$ is defined by
\begin{equation}
 \mathcal{P}^{\rm eq}_{\mathbf{d}_0}(\mathbf{l})=V_{\mathbf{00}}(\mathbf{l};
 \mathbf{d}_0) \equiv\int d\mathbf{Q}\Psi_{\mathbf{0}}^R(\mathbf{Q})\delta(\sum_{k=1}^N A_k\mathbf{q}_k + \mathbf{d}_0- \mathbf{l})\Psi_{\mathbf{0}}^L(\mathbf{Q})
 \label{non markovian pdf starting}
\end{equation}
Applying the Fourier transform 
$\tilde{f}(s)=\frac{1}{2\pi}\int_{-\infty}^\infty \rmd x
f(x)\ee{-isx}$ component-wise
to Eq.~(\ref{non markovian pdf starting}) (i.e. $\mathbf{l}\to\mathbf{s}$) yields
\begin{eqnarray}
 &\frac{1}{(2\pi)^3} \int d\mathbf{Q} \prod_{k=1}^N \left(\frac{\mu_k}{2\pi}\right)^{3/2}\exp\left(-\sum_{k=1}^N \frac{\mu_k}{2} \mathbf{q}^2_k + i(A_k\mathbf{q}_k + \mathbf{d}_0)\mathbf{s}\right) =\\
  &\frac{1}{(2\pi)^3}\mathrm{e}^{-s^2\sum_{k=1}^NA_k^2/2\mu_k+i \mathbf{d}_0\mathbf{s}}.
\end{eqnarray}
Inverting the Fourier transform we obtain, defining $\eta_0 = \sum_{k=1}^N A_k^2/(2\mu_k)$,
\begin{equation}
 \mathcal{P}^{\rm eq}_{\mathbf{d}_0}(\mathbf{l})=V_{\mathbf{00}}(\mathbf{l};\mathbf{d}_0)=\frac{1}{(2\pi)^3}\left(\frac{\pi}{\eta_0}\right)^{3/2}\mathrm{e}^{-(\mathbf{l}-\mathbf{d}_0)^2/4\eta_0}.
\end{equation}
Since we are only interested in the distance and not the direction we
need to marginalize over angles, i.e. 
\begin{equation}
 \int_0^\infty \rmd x x^2 \int_0^{2\pi} \rmd\phi \int_{-1}^1 \rmd(\cos\theta) V_{\mathbf{00}}(\mathbf{x}) \delta(|\mathbf{x}|-x),
\end{equation}
where $\phi$ is the polar angle, $\theta$ is the azimuthal angle and
without loss of generality we choose a frame of reference such that the vector $\mathbf{d}_0$ is parallel to the $z$-axis. The solution of this integral finally gives Eq~(\ref{non markovian pdf}):
\begin{equation}
 \mathcal{P}^{\rm eq}_{d_0}(l)=V_{\mathbf{00}}(l; d_0)=\frac{1}{\sqrt{\pi\eta_0}}\frac{l}{d_0}\mathrm{e}^{-(l^2+d_0^2)/4\eta_0}\sinh\left(\frac{ld_0}{2\eta_0}\right).
\end{equation}

\section{Spectral solution for $\mathcal{G}_{d_0}$}
\label{appendix:green series}
In the spectral solution for the Green's function in Eq.~(\ref{green
  series}) we have defined the elements
$V_{\mathbf{0N}}(l;d_0),V_{\mathbf{N0}}(l;d_0)$ which are derived as
follows. Let
\begin{eqnarray}
    &V_{\mathbf{0N}}(\mathbf{l};\mathbf{d}_0) =\int d\mathbf{Q}\Psi_{\mathbf{N}}^R(\mathbf{Q})\delta(\sum_{k=1}^N A_k\mathbf{q}_k + \mathbf{d}_0- \mathbf{l})\Psi_{\mathbf{0}}^L(\mathbf{Q}),\\
    &V_{\mathbf{N0}}(\mathbf{l};\mathbf{d}_0) =\int d\mathbf{Q}\Psi_{\mathbf{0}}^R(\mathbf{Q})\delta(\sum_{k=1}^N A_k\mathbf{q}_k + \mathbf{d}_0- \mathbf{l})\Psi_{\mathbf{N}}^L(\mathbf{Q}).
\end{eqnarray}
Fortunately, the above elements  $V_{\mathbf{0N}}$ and $V_{\mathbf{N0}}$ are identical (cf. Eq~(\ref{many body eigenfunctions})).
Therefore what we need to solve for is
\begin{eqnarray}
 &V_{\mathbf{N0}}(\mathbf{l};\mathbf{d_0})=\prod_{k=1}^N \int d\mathbf{q}_k \left(\frac{\mu_k}{2\pi}\right)^\frac{3}{2}\sqrt{\frac{1}{2^{n_{kx}+n_{ky}+n_{kz}}n_{kx}!n_{ky}!n_{kz}!}}\times\nonumber\nonumber\\
 &H_{n_{kx}}\left(\sqrt{\frac{\mu_k}{2}} q^x_k\right)H_{n_{ky}}\left(\sqrt{\frac{\mu_k}{2}} q^y_k\right)H_{n_{kz}}\left(\sqrt{\frac{\mu_k}{2}}q^z_k\right) \times\nonumber \nonumber\\
 &\ee{-\mu_k\mathbf{q}_k^2/2}\delta\left(\sum_{k=1}^N A_k \mathbf{q}_k+\mathbf{d}_0-\mathbf{l}\right).
 \label{higher order overlaps}
\end{eqnarray}
It is convenient to define the auxiliary variables
$\{\mathbf{q}'_k\}\equiv\{q^x_k-d^x_{0},q^y_k-d^y_{0},q^z_k-d^z_{0}\}$,
and then perform the Fourier transform $\mathbf{l}\to\mathbf{s}$ to obtain
\begin{eqnarray}
  &\frac{1}{(2\pi)^3}\prod_{k=1}^N \int d\mathbf{q}'_k \left(\frac{\mu_k}{2\pi}\right)^\frac{3}{2}\sqrt{\frac{1}{2^{n_{kx}+n_{ky}+n_{kz}}n_{kx}!n_{ky}!n_{kz}!}}\times\nonumber \nonumber \\
  &H_{n_{kx}}\left(\sqrt{\frac{\mu_k}{2}} q^{'x}_k\right)H_{n_{ky}}\left(\sqrt{\frac{\mu_k}{2}}q^{'x}_k\right)H_{n_{kz}}\left(\sqrt{\frac{\mu_k}{2}}q^{'z}_k\right)\nonumber\\
  &\ee{-\mu_k\mathbf{q}_k^{'2}/2-i A_k\mathbf{s}\cdot\mathbf{q}'_k}.
\end{eqnarray}
Factorizing in the three spatial dimensions, completing the square in
the exponential, and changing the variable to $t^{h}_k=\sqrt{\mu_k}
q^{'h}_{k}/\sqrt{2}$ (where the subscript $h$ denotes the respective
spatial coordinate) we find
\begin{eqnarray}
 \frac{1}{(2\pi)^3}\prod_{k=1}^N \left(\frac{1}{\pi}\right)^\frac{3}{2}\prod_{h=1}^3\sqrt{\frac{1}{2^{n_{kh}}n_{kh}!}}\mathrm{e}^{-s_h^2(A_k)^2/2\mu_k}\times\nonumber \nonumber\\
 \int_{-\infty}^\infty d t^h_{k} H_{n_{kh}}(t^h_{k})\exp\left(-\left[t^h_{k}-\left(-\frac{iA_k}{\sqrt{2\mu_k}}s_h\right)\right]^2\right)
\end{eqnarray}
whose solution is~\cite{gradshteyn_i_s_and_ryzhik_i_m_table_2007}
\begin{equation}
  \frac{1}{(2\pi)^3}\prod_{k=1}^N\prod_{h=1}^3\sqrt{\frac{2^{n_{kh}}}{n_{kh}!}}\left(-\frac{iA_k}{\sqrt{2\mu_k}}s_h\right)^{n_{kh}}\ee{-s_h^2(A_k)^2/2\mu_k}.
\label{blah}  
\end{equation}
It turns out to be convenient to write Eq.~(\ref{blah}) as
\begin{equation}
 \frac{1}{(2\pi)^3}\left[\prod_{k=1}^N\prod_{h=1}^3\sqrt{\frac{2^{n_{kh}}}{n_{kh}!}}\left(-\frac{iA_k}{\sqrt{2\mu_k}}\right)^{n_{kh}}\right]\prod_{h=1}^3 s_h^{\sum_{k=1}^N n_{kh}}\ee{-s_h^2\sum_{k=1}^N(A_k)^2/2\mu_k},
\end{equation}
and to define\footnote{We use the convention $0^0=1$ for terms $\left(\frac{A_k}{\sqrt{2\mu_k}}\right)^{n_{kx}+n_{ky}+n_{kz}}$.} 
\begin{eqnarray}
&M\equiv\prod_{k=1}^N\prod_{h=1}^3\sqrt{\frac{2^{n_{kh}}}{n_{kh}!}}\left(-\frac{iA_k}{2\mu_k}\right)^{n_{kh}}= \nonumber \\
  &\sqrt{\frac{2^{\sum_{k=1}^N\sum_{h=1}^3n_{kh}}}{\prod_{k=1}^N\prod_{h=1}^3n_{kh}!}}(-i)^{\sum_{k=1}^N\sum_{h=1}^3n_{kh}}\prod_{k=1}^N\left(\frac{A_k}{\sqrt{2\mu_k}}\right)^{n_{kx}+n_{ky}+n_{kz}}.
  \label{MM}
\end{eqnarray}
We now invert the Fourier transform
\begin{equation}
 \frac{M}{(2\pi)^3}\prod_{h=1}^3\int_{-\infty}^\infty d s_h s_h^{\sum_{k=1}^N n_{kh}}\mathrm{e}^{-\eta_0 s_h^2+is_hd_h}.
\end{equation}
Completing the square in the exponential and defining $t_h=\sqrt{\eta_0}s_h$, we can write:
\begin{equation}
 \frac{M}{(2\pi)^3}\prod_{h=1}^3\frac{\mathrm{e}^{-d_h^2/4\eta_0}}{\sqrt{\eta_0}^{\sum_{k=1}^N
       n_{kh} +1}}\int_{-\infty}^\infty d t_h t_h^{\sum_{k=1}^N
     n_{kh}}\mathrm{e}^{-(t_h-id_h/2\sqrt{\eta_0})^2},
\end{equation}
the integral in the previous equation can be solved analytically~\cite{gradshteyn_i_s_and_ryzhik_i_m_table_2007}
\begin{equation}
 \frac{M}{(2\pi)^3}\prod_{h=1}^3\frac{\mathrm{e}^{-d_h^2/4\eta_0}}{\sqrt{\eta_0}^{\sum_{k=1}^N n_{kh} +1}}\sqrt{\pi}(2i)^{-\sum_{k=1}^N n_{kh}}(-1)^{\sum_{k=1}^N n_{kh}} H_{\sum_{k=1}^N n_{kh}}\left(\frac{l_h}{2\sqrt{\eta_0}}\right).
\end{equation}
Using the definition of $M$ in Eq.~(\ref{MM}), defining
$N_h=\sum_{k=1}^N n_{kh}$ and $\mathcal{N}=N_x+N_y+N_z$, and going
back to the original, non-shifted coordinates we arrive at the following
form of Eq.~(\ref{higher order overlaps})
\begin{eqnarray}
 V_{\mathbf{0N}}(\mathbf{l};\mathbf{d}_0)=\frac{1}{(2\sqrt{\pi})^3}\sqrt{\frac{1}{2^\mathcal{N}\prod_{k=1}^Nn_{kx}!n_{ky}!n_{kz}!}}\prod_{k=1}^N\left(\frac{A_k}{\sqrt{2\mu_k}}\right)^{n_{kx}+n_{ky}+n_{kz}}\times\nonumber \nonumber \\
 \frac{1}{\sqrt{\eta_0}^{\mathcal{N}+3}} H_{N_x}\left(\frac{l^x-d^x_{0}}{2\sqrt{\eta_0}}\right)\!H_{N_y}\left(\frac{l^y-d^y_{0}}{2\sqrt{\eta_0}}\right)\!H_{N_z}\left(\frac{l^z-d^z_{0}}{2\sqrt{\eta_0}}\right)\mathrm{e}^{-(\mathbf{l}-\mathbf{d}_0)^2/4\eta_0}\!.
\end{eqnarray}
To integrate over the angular part it is convenient to use the following expansion of the Hermite polynomials~\cite{abramowitz_handbook_2013}:
\begin{equation}
   H_n(x)=n!\sum_{m=0}^{\lfloor\frac{n}{2}\rfloor}\frac{(-1)^m}{m! (n-2m)!}(2x)^{n-2m}.
\end{equation}
If we rotate our frame of reference such that $\hat{z} \parallel d^z_{0}$, (i.e.: $d^x_{0}=0$, $d^y_{0}=0$ and $d^z_{0}=d_0$) we find
 \begin{eqnarray}
 V_{\mathbf{0N}}(\mathbf{l};\mathbf{d}_0)=\frac{1}{(2\sqrt{\pi})^3}\sqrt{\frac{1}{2^\mathcal{N}\prod_{k=1}^Nn_{kx}!n_{ky}!n_{kz}!}}\prod_{k=1}^N\left(\frac{A_k}{\sqrt{2\mu_k}}\right)^{n_{kx}+n_{ky}+n_{kz}}\times\nonumber\nonumber\\
 \frac{1}{\sqrt{\eta_0}^{\mathcal{N}+3}} N_x! N_y! N_z! \mathrm{e}^{-\frac{d^2+d_0^2}{4\eta_0}}\times\nonumber\nonumber\\
 \sum_{a=0}^{\lfloor N_x/2\rfloor}\sum_{b=0}^{\lfloor N_y/2\rfloor}\sum_{c=0}^{\lfloor N_z/2\rfloor}\frac{(-1)^{a+b+c}}{a!b!c!(N_x-2a)!(N_y-2b)!(N_z-2c)!}\times\nonumber\nonumber\\
 \left(\frac{l^x}{\sqrt{\eta_0}}\right)^{N_x-2a}\left(\frac{l^y}{\sqrt{\eta_0}}\right)^{N_y-2b}\left(\frac{l^z-d_0}{\sqrt{\eta_0}}\right)^{N_z-2c}\mathrm{e}^{\mathbf{l}\cdot\mathbf{d}_0/2\eta_0}.
\end{eqnarray}
Using the binomial expansion of $(l^z-d_0)/\sqrt{\eta_0})^{N_z-2c}$ we obtain
\begin{eqnarray}
 V_{\mathbf{0N}}(\mathbf{l};\mathbf{d}_0)=\frac{1}{(2\sqrt{\pi})^3}\sqrt{\frac{1}{2^\mathcal{N}\prod_{k=1}^Nn_{kx}!n_{ky}!n_{kz}!}}\prod_{k=1}^N\left(\frac{A_k}{\sqrt{2\mu_k}}\right)^{n_{kx}+n_{ky}+n_{kz}}\times\nonumber\\
 \frac{1}{\sqrt{\eta_0}^{\mathcal{N}+3}} N_x! N_y! N_z! \mathrm{e}^{-(l^2+d_0^2)/4\eta_0}\times\nonumber\\
 \sum_{a=0}^{\lfloor N_x/2\rfloor}\sum_{b=0}^{\lfloor
   N_y/2\rfloor}\sum_{c=0}^{\lfloor
   N_z/2\rfloor}\frac{(-1)^{a+b+c}}{a!b!c!(N_x-2a)!(N_y-2b)!}\left(\frac{l^x}{\sqrt{\eta_0}}\right)^{N_x-2a}\left(\frac{l^y}{\sqrt{\eta_0}}\right)^{N_y-2b}\mathrm{e}^{\mathbf{l}\cdot\mathbf{d}_0/2\eta_0}\times\nonumber\\
 \sum_{m=0}^{N_z-2c} \frac{1}{m!(N_z-2c-m)!}\left(\frac{l^z}{\sqrt{\eta_0}}\right)^{N_z-2c-m}\left(-\frac{d_0}{\sqrt{\eta_0}}\right)^{m}.
\end{eqnarray}
At this point we can integrate over the angles of $\mathbf{l}$ fixing
the length, hence 
\begin{eqnarray}
 V_{\mathbf{0N}}(l,d_0)=\frac{1}{(2\sqrt{\pi})^3}\sqrt{\frac{1}{2^\mathcal{N}\prod_{k=1}^Nn_{kx}!n_{ky}!n_{kz}!}}\prod_{k=1}^N\left(\frac{A_k}{2_k}\right)^{n_{kx}+n_{ky}+n_{kz}}\times\nonumber\nonumber \\
 \frac{1}{\eta_0^{\mathcal{N}+1}} N_x! N_y! N_z! \mathrm{e}^{-(l^2+d_0^2)/4\eta_0^2}\times\nonumber\\
 \sum_{a=0}^{\lfloor N_x/2\rfloor}\sum_{b=0}^{\lfloor N_y/2\rfloor}\sum_{c=0}^{\lfloor N_z/2\rfloor}\frac{(-1)^{a+b+c}}{a!b!c!(N_x-2a)!(N_y-2b)!}\left(\frac{l}{\eta_0}\right)^{N_x+N_y-2(a+b)+2}\times\nonumber\\
 \int_0^{2\pi} d\phi (\sin \phi)^{N_y-2b} (\cos \phi)^{N_x-2a}\times\nonumber\\
 \sum_{m=0}^{N_z-2c} \frac{1}{m!(N_z-2c-m)!}\left(\frac{l}{\eta_0}\right)^{N_z-2c-m}\left(-\frac{d_0}{\eta_0}\right)^{m}\times\nonumber \\
 \int_0^{\pi} d\theta
 (\sin\theta)^{N_x+N_y-2(a+b)+1}(\cos\theta)^{N_z-2c-m}\mathrm{e}^{\cos\theta
   ld_0/2\eta_0^2}.
\end{eqnarray}
The first integral is 
\begin{equation}
 \int_0^{2\pi} d\phi \cos^{n}\phi\sin^{m}\phi=\frac{\pi n!m!}{2^{n+m-1}(\frac{n}{2})!(\frac{m}{2})!(\frac{n+m}{2})!} ,
\end{equation}
and is non-zero only if $n$ and $m$ are even~\cite{wolfram_research_inc_mathematica_2019}. Therefore $N_x$ and $N_x$ must be even.
While the second integral reads~\cite{wolfram_research_inc_mathematica_2019}
\begin{eqnarray}
 \int_0^{\pi} d\theta (\sin \theta)^{n}(\cos\theta)^{m}\mathrm{e}^{k\cos\theta} =\nonumber\\
 \frac{\sqrt{\pi}}{4}\gamma\left(\frac{1+n}{2}\right)\left[2(1+(-1)^m)\gamma\left(\frac{1+m}{2}\right)\, _1\tilde{F}_2\left(\frac{1+m}{2};\frac{1}{2},\frac{2+m+n}{2};\frac{k^2}{4}\right) \right.\nonumber\\
 \left.-(-1+(-1)^m) k \gamma\left(1+\frac{m}{2}\right)\, _1\tilde{F}_2\left(\frac{2+m}{2};\frac{3}{2},\frac{3+m+n}{2};\frac{k^2}{4}\right) \right],
\end{eqnarray}
where we have introduced the Euler's gamma function $\gamma(x)$ as
well as the regularized hypergeometric function $_p\tilde{F}_q(a_1,\cdots,a_p;b_1,\cdots,b_q;x)$~\cite{abramowitz_handbook_2013}.
Putting all together we finally arrive at
\begin{eqnarray}
 V_{\mathbf{0N}}(l;d_0)=\frac{1}{16}\sqrt{\frac{1}{2^\mathcal{N}\prod_{k=1}^Nn_{kx}!n_{ky}!n_{kz}!}}\prod_{k=1}^N\left(\frac{A_k}{\sqrt{2\mu_k}}\right)^{n_{kx}+n_{ky}+n_{kz}}\times\nonumber\\
 \frac{1}{\sqrt{\eta_0}^{\mathcal{N}+1}} N_x! N_y! N_z! \mathrm{e}^{-(d^2+d_0^2)/4\eta_0}\times\nonumber\\
 \sum_{a=0}^{N_x/2}\sum_{b=0}^{ N_y/2}\sum_{c=0}^{\lfloor N_z/2\rfloor}\frac{(-1)^{a+b+c}}{a!b!c!(\frac{N_x-2a}{2})!\left(\frac{N_y-2b}{2}\right)!2^{N_x+N_y-2(a+b)}} \times\nonumber\\
 \sum_{m=0}^{N_z-2c} \frac{1}{m!(N_z-2c-m)!}\left(\frac{l}{\sqrt{\eta_0}}\right)^{\mathcal{N}-2(a+b+c)-m+2}\left(-\frac{d_0}{\sqrt{\eta_0}}\right)^{m}\times\nonumber \\
\left[2(1+(-1)^{N_z-2c-m})\gamma\left(\frac{1+N_z-2c-m}{2}\right)\right.\times\nonumber \\ _1\tilde{F}_2\left(\frac{1+N_z-2c-m}{2};\frac{1}{2},\frac{3+\mathcal{N}-2(a+b+c)-m}{2};\frac{l^2d_0^2}{16\eta_0^2}\right)\nonumber\\
 -(-1+(-1)^{N_z-2c-m}) \frac{ld_0}{2\eta_0} \gamma\left(1+\frac{N_z-2c-m}{2}\right)\times\nonumber \\
 \left._1\tilde{F}_2\left(\frac{2+N_z-2c-m}{2};\frac{3}{2},\frac{4+\mathcal{N}-2(a+b+c)-m}{2};\frac{l^2d_0^2}{16\eta_0^2}\right) \right];
 \label{final V}
\end{eqnarray}
which finally allows us to write down the non-Markovian Green's function expressed as an infinite series in Eq.~\ref{green series}.
In addition, the series expansion allows us the compute the cross conditioned Green's function
\begin{equation}
    \mathcal{G}_{d_0,d_0^\prime}(l,t|l')=V_{\mathbf{00}}(l';d_0^\prime)^{-1}\sum_{\mathbf{N}}V_{\mathbf{0N}}(l;d_0)V_{\mathbf{N0}}(l';d_0^\prime)\ee{-\Lambda_Nt}
    \label{green series cross}
\end{equation}
that is the probability that the distance between the beads $i$ and
$j$ is equal to $l$ at time $t$ conditioned to the fact that the
distance between the beads $k$ and $l$ at time $0$ was equal to $l'$,
assuming that these two distances have rest lengths $d_0$ and
$d_0^\prime$, respectively. In particular in order to evaluate $V_{\mathbf{N0}}(l';d_0^\prime)$ we need to consider that the distance $\mathbf{l}'$ is expressed via the normal coordinates as
\begin{equation}
    \mathbf{d}'=\mathbf{r}_k-\mathbf{r}_l=\sum_{i=1}^N B_i\mathbf{q}_i,
\end{equation}
and we in turn use these coefficients to define $\zeta_t=\sum_{k=1}^N
B_k^2/2\mu_k\ee{-\mu_kt}$ and $\zeta_0=\sum_{k=1}^N
B_k^2/2\mu_k$ instead of $\eta_t$ and $\eta_0$.

\section{Closed form solution for $\mathcal{G}_{d_0}$}
\label{appendix: green closed}
In order to obtain the equivalent result in a closed form solution we should consider the following integral:
\begin{eqnarray}
 \mathcal{J}_{\mathbf{d}_0}(\mathbf{l},t;\mathbf{l}_1)=\int \rmd\mathbf{Q}\int \rmd\mathbf{Q}_1 G(\mathbf{Q},t|\mathbf{Q}_1) P_\mathrm{eq}(\mathbf{Q}_1)\times\nonumber\\
 \delta(\sum_{k=1}^N A_k \mathbf{q}_{1k}+\mathbf{d}_{0}-\mathbf{l}_1)\delta(\sum_{k=1}^N A_k \mathbf{q}_k+\mathbf{d}_0-\mathbf{l}).
\end{eqnarray}
Performing the first Fourier transform, between
$\mathbf{l}_1\to\mathbf{u}$ the above integral becomes
\begin{eqnarray}
 \int d\mathbf{Q}\delta(\sum_{k=1}^N A_k \mathbf{q}_k+\mathbf{d}_0-\mathbf{l})\ee{-i\mathbf{d_0}\cdot\mathbf{u}}\times\nonumber\\
 \left(\frac{1}{2\pi}\right)^3 \int d\mathbf{q}_{1k} \prod_{k=1}^N\left(\frac{\mu_k^2}{2\pi}\right)^{3/2}\left(\frac{\mu_k}{2\pi(1-\mathrm{e}^{-2\mu_k t})}\right)^{3/2} \times\nonumber\\
 \exp\left[-\frac{\mu_k}{2(1-\mathrm{e}^{-2\mu_k^2t})}\left(\mathbf{q}_k^2+\mathbf{q}_{1k}^2\mathrm{e}^{-2\mu_kt}-2\mathbf{q}_k\cdot\mathbf{q}_{1k} \mathrm{e}^{-\mu_k t}\right)\right]\times \nonumber\\ \mathrm{e}^{-i A_k \mathbf{q}_{1k}\cdot\mathbf{u}} \mathrm{e}^{-\mu_k\mathbf{q}_{1k}^2/2};
\end{eqnarray}
and the integration yields
\begin{eqnarray}
 \left(\frac{1}{2\pi}\right)^3\int d\mathbf{Q}\delta(\sum_{k=1}^N A_k \mathbf{q}+\mathbf{d}_0-\mathbf{l})\ee{-i\mathbf{d_0}\cdot\mathbf{u}}\times\nonumber\\ \prod_{k=1}^N\left(\frac{\mu_k^2}{2\pi}\right)^{3/2}
 \exp\left[-\frac{\mu_k}{2}\left(\mathbf{q}_k^2+2i\frac{A_k}{\mu_k}\mathrm{e}^{-\mu_k t}\mathbf{u}\cdot\mathbf{q}_k + \frac{A_k^2}{\mu_k^2}(1-\mathrm{e}^{-2\mu_kt})\mathbf{u}^2\right)\right].
\end{eqnarray}
Performing the second Fourier transform $\mathbf{l}\to\mathbf{v}$ we find
\begin{eqnarray}
  \ee{-i\mathbf{d}_0\cdot(\mathbf{u}+\mathbf{v})}\left(\frac{1}{2\pi}\right)^6\prod_{k=1}^N\int \rmd\mathbf{q}_k \left(\frac{\mu_k}{2\pi}\right)^{3/2} \times\nonumber\\
  \exp\left[-\frac{\mu_k}{2}\left(\mathbf{q}_k^2+2i\frac{A_k}{\mu_k}(\mathrm{e}^{-\mu_kt}\mathbf{u}+\mathbf{v})\cdot\mathbf{q}_k+\frac{A_k^2}{\mu_k^2}(1-\mathrm{e}^{-2\mu_kt})\mathbf{u}^2\right)\right],
\end{eqnarray}
that reads
\begin{equation}
 \ee{-i\mathbf{d}_0\cdot(\mathbf{u}+\mathbf{v})}(\frac{1}{2\pi})^6\prod_{k=1}^N \exp\left[-\frac{A_k^2}{2\mu_k}(\mathbf{u}^2+\mathbf{v}^2)+2\frac{A_k^2}{2\mu_k}\mathrm{e}^{-\mu_k t}\mathbf{u}\cdot\mathbf{v}\right].
\end{equation}
It is convenient to define
\begin{equation}
\sum_{k=1}^N \frac{A_k^2}{2\mu_k}\mathrm{e}^{-\mu_kt}=\eta_t\,\to \sum_{k=1}^N \frac{A_k^2}{2\mu_k}=\eta_0
\end{equation}
so the Fourier transform of the joint-density is:
\begin{equation}
 \tilde{\mathcal{J}}_{\mathbf{d}_0}(\mathbf{v},t;\mathbf{u})=\frac{1}{(2\pi)^6} \exp\left(-\eta_0 \mathbf{u}^2 - \eta_0 \mathbf{v}^2 + 2 \eta_t \mathbf{u}\cdot\mathbf{v} + i \mathbf{d_0}\cdot(\mathbf{u}+\mathbf{v})\right).
\end{equation} 
The inversion of the two Fourier transforms gives straightforwardly
\begin{eqnarray}
 \mathcal{J}_{\mathbf{d}_0}(\mathbf{l},t;\mathbf{l}_1)=\frac{1}{2^6\pi^3}\left(\frac{1}{\eta_0^2-\eta_t^2}\right)^{3/2} \times\nonumber\\
  \exp\left[-\frac{\eta_0(\mathbf{l}-\mathbf{d_0})^2+\eta_0(\mathbf{l}_1-\mathbf{d_0})^2-2\eta_t(\mathbf{l}-\mathbf{d_0})\cdot(\mathbf{l}_1-\mathbf{d}_0)}{4(\eta_0^2-\eta_t^2)}
  \right],
\end{eqnarray}
We now marginalize over the angles
\begin{equation}
  \mathcal{J}_{d_0}(l,t;l_1)\equiv \int d\mathbf{d} \int d\mathbf{d}_1\int d\mathbf{d}_0 \delta(|\mathbf{d}_0|-d_0) \delta(|\mathbf{l}_1|-l_1) \delta(|\mathbf{l}|-l)\mathcal{J}_{\mathbf{d}_0}(\mathbf{l}, t ; \mathbf{l}_1),
\end{equation}
by moving to a frame of reference where $\mathbf{d}_0$ is parallel to
the the $z$ axis, and express all the vectors in spherical
coordinates. This removes all delta-functions and $d_0$ in the new
frame of reference is just a scalar. By doing so we obtain
\begin{eqnarray}
 \mathcal{J}_{d_0}(l,t;l_1)=\frac{1}{2^6\pi^3}\left(\frac{1}{\eta_0^2-\eta_t^2}\right)^{3/2}\exp\left(-\frac{\eta_0 l^2+\eta_0 l_1^2+2(\eta_0-\eta_t)d_0^2}{4(\eta_0^2-\eta_t^2)}\right)l^2l_1^2\times\nonumber\\
 \int_0^{2\pi} d\phi \int_0^{2\pi} d\phi' \int_{-1}^{1}d(\cos\theta)\int_0^\pi d(\cos\theta')\times\nonumber\\
 \exp\left[\frac{(\eta_0-\eta_t)ld_0}{2(\eta_0^2-\eta_t^2)}\cos\theta+\frac{(\eta_0-\eta_t)l_1d_0}{2(\eta_0^2-\eta_t^2)}\cos\theta'+\frac{\eta_t ll_1}{2(\eta_0^2-\eta_t^2)}\right.\times\nonumber\\
 \left.(\cos\phi\cos\phi'\sin\theta\sin\theta'+\sin\phi\sin\phi'\sin\theta\sin\theta'+\cos\theta\cos\theta')\Bigg]\right..
\end{eqnarray}
The two integrals over $\phi$ and $\phi'$ (keeping in mind that $\cos(\phi-\phi')=\cos\phi\cos\phi'+\sin\phi\sin\phi'$) give us
\begin{eqnarray}
 \frac{1}{16 \pi}\left(\frac{1}{\eta_0^2-\eta_t^2}\right)^{3/2}\exp\left(-\frac{\eta_0 l^2+\eta_0 l_1^2+2(\eta_0-\eta_t)d_0^2}{4(\eta_0^2-\eta_t^2)}\right)l^2l_1^2 \times\nonumber\\
 \int_{-1}^{1}d(\cos\theta)\int_0^\pi d(\cos\theta') \exp\left[\frac{(\eta_0-\eta_t)ld_0}{2(\eta_0^2-\eta_t^2)}\cos\theta+\frac{(\eta_0-\eta_t)l_1d_0}{2(\eta_0^2-\eta_t^2)}\cos\theta'\right.\nonumber\\ 
 \left.+\frac{\eta_t ll_1}{2(\eta_0^2-\eta_t^2)}\cos\theta\cos\theta'\right]
 I_0\left(\frac{\eta_t ll_1}{2(\eta_0^2-\eta_t^2)}\sqrt{1-\cos^2\theta}\sqrt{1-\cos^2\theta'}\right),
\end{eqnarray}
where $I_0(x)$ is the modified Bessel function of the first kind.
The first integral in $\cos\theta'$ is
solvable~\cite{gradshteyn_i_s_and_ryzhik_i_m_table_2007}, and by changing the variable $\cos\theta \to x$ we are left with
\begin{eqnarray}
 \frac{1}{8 \pi}\left(\frac{1}{\eta_0^2-\eta_t^2}\right)^{3/2}\exp\left(-\frac{\eta_0 l^2+\eta_0 l_1^2+2(\eta_0-\eta_t)d_0^2}{4(\eta_0^2-\eta_t^2)}\right)l^2l_1^2 \times\nonumber\\
 \int_{-1}^{1}dx \mathrm{e}^{\frac{(\eta_0-\eta_t)ld_0}{2(\eta_0^2-\eta_t^2)}x}\frac{\sinh\left(\sqrt{\frac{(\eta_0-\eta_t)^2l_1^2d_0^2+\eta_t^2l^2l_1^2+2\eta_t(\eta_0-\eta_t)ll_1^2d_0 x}{4(\eta_0^2-\eta_t^2)^2}}\right)}{\sqrt{\frac{(\eta_0-\eta_t)^2l_1^2d_0^2+\eta_t^2l^2l_1^2+2\eta_t(\eta_0-\eta_t)ll_1^2d_0 x}{4(\eta_0^2-\eta_t^2)^2}}}.
 \label{with numerical integral}
\end{eqnarray}
And the final integral yields~\cite{wolfram_research_inc_mathematica_2019} 
\begin{eqnarray}
 \mathcal{J}_{d_0}(l,t;l_1)=\frac{1}{16\sqrt{\pi}}\left(\frac{1}{\eta_0^2-\eta_t^2}\right)^{3/2}\exp\left(-\frac{\eta_0 l^2+\eta_0 l_1^2+2(\eta_0-\eta_t)d_0^2}{4(\eta_0^2-\eta_t^2)}\right)\times\nonumber\\
l^2l_1^2\frac{\mathrm{e}^{-ab/c-c/4a}}{\sqrt{ac}}  \left[\mathrm{erfi}\left(\frac{2a\sqrt{b-c}-c}{2\sqrt{ac}}\right)-\mathrm{erfi}\left(\frac{2a\sqrt{b-c}+c}{2\sqrt{ac}}\right)+\right.\nonumber\\
 \left.\mathrm{erfi}\left(\frac{c-2a\sqrt{b+c}}{2\sqrt{ac}}\right)+\mathrm{erfi}\left(\frac{c+2a\sqrt{b+c}}{2\sqrt{ac}}\right)\right]
\end{eqnarray}
having defined
\begin{eqnarray}
 & a =\frac{(\eta_0-\eta_t)ld_0}{2(\eta_0^2-\eta_t^2)}, \\
 &b=\frac{(\eta_0-\eta_t)^2l_1^2d_0^2+\eta_t^2l^2l_1^2}{4(\eta_0^2-\eta_t^2)^2}, \\
 &c=\frac{\eta_t(\eta_0-\eta_t)ll_1^2d_0}{2(\eta_0^2-\eta_t^2)^2};
\end{eqnarray}
the direct substitution of these auxiliary variables gives, upon
division by $\mathcal{P}^{\rm eq}_{d_0}$ and some
simplification, Eq.~(\ref{joint closed}).

\section{Derivation of equilibrium autocorrelation function}
\label{appendix: autocorrelation}
In order to compute the autocorrelation function in Eq.~(\ref{elem2}) the following integrals must be evaluated
\begin{equation}
    \mathcal{V}^{d_0}_{\mathbf{0N}}=\int_0^\infty d x  V_{\mathbf{0N}}(x,d_0)x,\quad \mathcal{V}^{d_0}_{\mathbf{N0}}=\int_0^\infty dx  V_{\mathbf{N0}}(x,d_0)x.
\end{equation}
These two integrals are identical and the integration yields~\cite{gradshteyn_i_s_and_ryzhik_i_m_table_2007}
\begin{eqnarray}
  \fl
 \mathcal{V}^{d_0}_{\mathbf{0N}}=
 \frac{1}{16}\sqrt{\frac{1}{2^\mathcal{N}\prod_{k=1}^Nn_{kx}!n_{ky}!n_{kz}!}}\prod_{k=1}^N\left(\frac{A_k}{\sqrt{2\mu_k}}\right)^{n_{kx}+n_{ky}+n_{kz}} \frac{1}{\sqrt{\eta_0}^{\mathcal{N}+1}} N_x! N_y! N_z! \mathrm{e}^{-d_0^2/4\eta_0}\times\nonumber\\
 \fl\sum_{a=0}^{N_x/2}\sum_{b=0}^{ N_y/2}\sum_{c=0}^{\lfloor N_z/2\rfloor}\!\frac{(-1)^{a+b+c}}{a!b!c!(\frac{N_x-2a}{2})!(\frac{N_y-2b}{2})!2^{N_x+N_y-2(a+b)}}
 \!\sum_{l=0}^{N_z-2c} \!\frac{1}{l!(N_z-2c-l)!}\left(-\frac{d_0}{\sqrt{\eta_0}}\right)^l\times\nonumber\\
\fl\left[(1+(-1)^{N_z-2c-l})\gamma\left(\frac{1+N_z-2c-l}{2}\right)2^
{\mathcal{N}-2(a+b+c)-l+4} \eta_0 \gamma\left(\frac{\mathcal{N}-2(a+b+c)-l+4}{2}\right) \right.\times\nonumber\\
\fl _2\tilde{F}_2\left(\frac{1+N_z-2c-l}{2},\frac{\mathcal{N}-2(a+b+c)-l+4}{2};\frac{1}{2},\frac{3+\mathcal{N}-2(a+b+c)-l}{2};\frac{d_0^2}{4\eta_0}\right)-\nonumber \\
\fl (-1+(-1)^{N_z-2c-l}) d_0\sqrt{\eta_0} \gamma\left(1+\frac{N_z-2c-l}{2}\right)\times\nonumber\\
 \fl 2^{\mathcal{N}-2(a+b+c)-l+3} \gamma\left(\frac{\mathcal{N}-2(a+b+c)-l+5}{2}\right)\times\nonumber\\
\fl \left._2\tilde{F}_2\left(\frac{2+N_z-2c-l}{2},\frac{\mathcal{N}-2(a+b+c)-l+5}{2};\frac{3}{2},\frac{4+\mathcal{N}-2(a+b+c)-l}{2};\frac{d_0^2}{4\eta_0}\right) \right]\nonumber\\
 \label{final D}
\end{eqnarray}
If we are instead interested in the cross-correlation the more general
Eq~(\ref{green series cross}) must be used and the two integrals differ
in therms of some constants, i.e. they are obtained by changing the
following variables $d_0\to d_0^\prime$, $\{A_k\}\to \{B_k\}$ and
$\eta_t \to\zeta_t$.

\subsection{Rouse-limit autocorrelation function}
In Fig.~\ref{fig:autocorr} we showed how the autocorrelation for a GNM
compares to the autocorrelation in the Rouse limit (i.e. $d_0\to
0$). The latter can be obtained in a closed form~\cite{lapolla_toolbox_2021}
\begin{eqnarray}
    &\mathcal{C}(t)=\frac{\langle l(t)l(0)\rangle- \langle l\rangle^2}{\langle l^2\rangle- \langle l\rangle^2};\\
    &\langle l(t)l(0)\rangle=\frac{4\left[3\eta_t\sqrt{\eta_0^2-\eta_t^2}+2(\eta_0^2+\eta_t^2)\arctan(\eta_t/(\eta_0^2-\eta_t^2))\right]}{\pi\eta_t},\nonumber\\
    \\
    &\langle l\rangle=4\sqrt{\eta_0/\pi},\quad
    \langle l^2\rangle=6\eta_0.
\end{eqnarray}

\section{Short-time expansion of $\mathcal{G}_{d_0}$}
\label{appendix:short time}
Introducing the auxiliary variable $\phi(t)=\eta_t/\eta_0$ in Eq.~(\ref{joint closed}) we can write the return joint-density as
and expanding to linear order in $t$ using 
\begin{eqnarray}
    &\phi(t) \stackrel{t\to0}{\simeq} 1-\frac{\sum_{k=1} A_k^2 t}{2\eta_0}
    &\phi^2(t) \stackrel{t\to0}{\simeq} 1-\sum_{k=1} A_k^2 \frac{t}{\eta_0};
    \label{expansion in t of phi}
\end{eqnarray}
we find the partial limits
\begin{eqnarray}
    &\exp\left(-\frac{2d^2\phi(t)+(1-\phi(t))d_0^2}{4\eta_0\phi(t)(1-\phi(t))}\right)\stackrel{t\to 0}{\sim} \ee{-1/t}\to 0,\\
    &\erfi\left({\frac{\pm2d\phi(t)+d_0(1-\phi(t))}{2\sqrt{\eta_0\phi(t)(1-\phi(t)^2)}}}\right)\stackrel{t\to 0}{\sim} \erfi(\pm t^{-1/2})\to \pm \infty,\\
    &\erfi\left({\frac{d_0(1-\phi(t))}{2\sqrt{\eta_0\phi(t)(1-\phi(t)^2)}}}\right)\stackrel{t\to 0}{\sim} \erfi(\sqrt{t})\to 0;
\end{eqnarray}
where all the convergences are of exponential order. Therefore, while
we can neglect the second $\erfi$, we need to retain the product between the exponential and the two diverging $\erfi$s and only then plug them into in Eq.~(\ref{expansion in t of phi}).
Thus considering the expansion for large and real arguments of
erfi~\cite{abramowitz_handbook_2013}
\begin{equation}
    \erfi(x)\stackrel{x\to \pm \infty, x\in\mathds{R}}{\simeq}\mp i+\left(\frac{1}{x}+\frac{1}{2x^3}+O(x^{-5})\right)\frac{\mathrm{e}^{x^2}}{\sqrt{\pi}},
\end{equation}
and explicitly, multiplying by the remaining exponentials
Eq.~(\ref{joint closed}) becomes (note that $\mathcal{P}_{d_0}^{\rm
  eq}(l)
\mathcal{G}_{d_0}(l,t|l)\equiv\mathcal{J}_{d_0}(l,t;l)$)
\begin{eqnarray}
\fl    \mathcal{J}_{d_0}(d,t;d)\stackrel{t\to 0}{\simeq}\frac{d^2}{8\pi
      d_0}\ee{-(l^2+d_0^2)/2\eta_0(1+\phi(t))}
    \times\nonumber\\
\fl    \Bigg\{\left[\frac{2\sqrt{1+\phi(t)}}{\sqrt{1-\phi(t)}\eta_0(-2d\phi(t)+d_0(1-\phi(t)))}+4\frac{\phi(t)\sqrt{1-\phi(t)}(1+\phi(t))^{3/2}}{(-2d\phi(t)+d_0(1-\phi(t)))^3}\right]\ee{-ld_0/\eta_0(1+\phi(t))}\nonumber\\
\fl    +\left[\frac{2\sqrt{1+\phi(t)}}{\sqrt{1-\phi(t)}\eta_0(2d\phi(t)+d_0(1-\phi(t)))}+4\frac{\phi(t)\sqrt{1-\phi(t)}(1+\phi(t))^{3/2}}{(2d\phi(t)+d_0(1-\phi(t)))^3}\right]\ee{ld_0/\eta_0(1+\phi(t))}\left.\right]\Bigg\}\nonumber\\.
\end{eqnarray}
Using Eq.~(\ref{expansion in t of phi}) and expanding $t=0$ and
introducing $\kappa=\sum_{k=1}^N A_k^2$ we finally
arrive at Eq.~(\ref{small}).

\section{Evaluation of the variance of the occupation time fraction}
\label{appendix:varianceLT calculation}
The direct implementation of Eq.~(\ref{local-time variance}) suffers
from slow convergence issues. We suspect that this problem has his
roots in the (well-known) slow convergence of series involving Hermite
polynomials~\cite{boyd_rate_1980}.
We therefore combine the analytical short-time asymptotics in
Eq.(\ref{SDEV}) with the spectral solution. Defining a small cutoff
time $t_s\ll 1$ and rewriting Eq.~(\ref{variance explict integral}) (using the linearity of integration) as
\begin{equation}
\fl
\sigma_{d_0}^2(l,t)=\frac{2\mathcal{P}^{\rm eq}_{d_0}(l)}{t}\int_0^{t_s}
d\tau(1-\tau/t)
\mathcal{G}_{d_0}(l,\tau|l)+\frac{2\mathcal{P}^{\rm eq}_{d_0}(l)}{t}\int_{t_s}^t
d\tau(1-\tau/t) [\mathcal{G}_{d_0}(l,\tau|l)-\mathcal{P}^{\rm eq}_{d_0}(l)].
\end{equation}
We can explicitly evaluate the first addend using Eq.~(\ref{SDEV}) and
evaluate the second
term using the spectral expansion (\ref{green series}). Note that the
first term in the series (with $\Lambda_\mathbf{0}=0$) must be treated
in a manner different thant the rest. Therefore $\sigma_{d_0}^2(l,t)$
can be conveniently written (and implemented) in the form
\begin{eqnarray}
  \sigma_{d_0}^2(d,t)=2\mathcal{P}^{\rm eq}_{d_0}(l)\left(\frac{8}{3\sqrt{\kappa\pi
    t}}+\frac{4}{15l^2}\sqrt{\frac{\kappa
        t}{\pi}}-\mathcal{P}^{\rm eq}_{d_0}(l)\right)\nonumber\\
    +\frac{2}{t^2}\sum_{\mathbf{N}\neq\mathbf{0}}V_{\mathbf{N0}}(l;d_0)V_{\mathbf{0N}}(l;d_0)\left[(t-t_s)\frac{\ee{-\Lambda_\mathbf{N} t_s}}{\Lambda_\mathbf{N}}-\frac{\ee{-\Lambda_\mathbf{N} t_s}-\ee{-\Lambda_\mathbf{N} t}}{\Lambda_\mathbf{N}^2}\right]\nonumber\\
    +\mathcal{P}_{d_0}^{\rm eq}(l)^2\left(\frac{t_s}{t}-2\right)\frac{t_s}{t}.
\end{eqnarray}

\section{Notes on the numerical implementation of the results}
\label{appendix:computer evaluation}
Accompanying this article there is a C++ implementation of all analytical
results. The code allows the computation the Green's function $\mathcal{G}_{d_0}$, the mean
$\langle \theta_t(l,d_0)\rangle$
and variance $\sigma_{d_0}^2(d,t)$ of the occupation time fraction, as well as the
autocorrelation function $\mathcal{C}_{d_0}(t)$ for a generic Gaussian
Network.
The connectivity matrix of the network $\Gamma$ must be provided as a
plain text file and is diagonalized using the
Armadillo libray~\cite{sanderson_armadillo_2016,
  sanderson_user-friendly_2018}.

A closed-form expression of the joint
density in Eq.~(\ref{joint closed}) is implemented in 
the available C++ code.
However, for numerical stability and speed of
computation it is convenient to implement Eq.~(\ref{with
  numerical integral}) and perform the final integral numerically using
a Gauss-Kronrod quadrature
routine~\cite{noauthor_httpswwwboostorgdoclibs1_73_0libsmathdochtmlmath_toolkitgauss_kronrodhtml_2020}.

Our main results are based on the evaluation of both, Eq.~(\ref{final
  V}) and Eq.~(\ref{final D}). Both require the evaluation of the less
common regularized hypergeometric functions $_p\tilde{F}_q$. A notable
exception is the Arblib library~\cite{johansson_arb_2017}, that
implements several "special" functions using arbitrary precision
arithmetic.  The reliable evaluation of such functions is challenging
and often requires several different methods to cover the entire
domain~\cite{johansson_computing_2019}.  Unfortunately this higher
reliability comes with a higher computational cost compared to machine
precision arithmetic.  However hypergeometric functions converge on
the entire complex plane if $p\leq
q$~\cite{johansson_computing_2019}. In addition, we only need to
evaluate them when all the parameters are positive real
numbers. Therefore we implemented the series definitions of these
function  directly since in our case these converge reasonably fast to
a desired accuracy as long as the parameters are not too large.

Many of our results, in particular the autocorrelation function and
the variance of the fraction of occupation time, can only be expressed
analytically using the eigendecomposition of the Fokker-Plank
operator.  Unfortunately the computational effort required in the
generation of all necessary terms to achieve convergence is huge. In
addition, this number scales non-polynomially with the number of beads
in the network. Therefore the attached program should be used with
care as it does not generate reliable results when the size of the network becomes too large.

\pagebreak
\bibliographystyle{unsrt}
\bibliography{elasticnetworkABB.bib}

\end{document}